\documentstyle[epsfig,12pt]{ioplppt}
\jl{6}

\begin{flushright}
\mbox{
\vbox{{\small \hbox{University of Parma UPRF 96-466}
              \hbox{gr-qc/9605064}
}}}
\end{flushright}

\begin{document}


\title{On the relation between the connection and \\
the loop representation of quantum gravity}[On the relation 
between the connection and the loop representation]

\author{Roberto De Pietri}
\address{Dipartimento di Fisica - Sezione di Fisica Teorica,
Universit\`a di Parma, 43100 Parma, Italy 
and  I.N.F.N., Sezione di Milano, Gruppo Collegato di Parma.\\
e-mail: depietri@vaxpr.pr.infn.it}


\begin{abstract}
Using Penrose's binor calculus for $SU(2)$ ($SL(2,C)$) tensor
expressions, a graphical method for the connection
representation of Euclidean Quantum Gravity (real connection)
is constructed.  It is explicitly shown that: 
{\it (i)} the recently proposed scalar product in the
loop-representation coincide with the Ashtekar-Lewandowski
cylindrical measure in the space of connections; 
{\it (ii)} it is possible to establish a correspondence 
between the operators in the connection representation and 
those in the loop representation.  
The construction is based on embedded spin
network, the Penrose's graphical method of $SU(2)$ calculus, and
the existence of a generalized measure on the space of
connections modulo gauge transformations.
\end{abstract}
\maketitle





\section{Introduction}

Recently progress has  been made  in the
Ashtekar\cite{AbhayVAR} formulation of Canonical Quantum
Gravity\cite{QG} and the relation between the two different
approach (the connection representation\cite{Ashtekar95a} and
the loop representation\cite{LOOP}) is becoming more
rigorous. The possibility of establishing such a relation relies
on the fact that both have a more elegant formulation in
terms of {\it spin-network} states
\cite{Ashtekar95a,Rovelli95,DePietri96}.  
However, the formulation in terms of {\it spin-networks} 
arises in the two representations in very different contexts.
In the loop-representation the  {\it spin-network} states
originate from the elimination of the so-called {\it
Mandelstam-relation}\cite{Rovelli95,DePietri96} while in
the connection representation they are the natural {\it
cylindrical-functions}\footnote{the term cylindrical
function comes from the language of Wiener integration on an
infinite-dimensional space.} in terms of which a
generalized measure\cite{FunctINT} on the space of connections
modulo gauge transformations ($\overline{{\cal A}/{\cal G}}$) 
is defined. It is worthwhile to recall that
the lattice regularization for canonical Quantum Gravity
proposed by R.\ Loll \cite{LollRETICOLO} amounts to consider 
graphs $\gamma$ that constitute a finite
cubic lattice.
The idea that it is possible to compare
operators defined in the loop and in the connection representation
was analyzed by J.\ Lewandowski in \cite{Lewandowski96}.

The main purpose of this article is to make a further step
in the direction of proving that the loop and the connection
representation are unitarily isomorphic. In particular,
we will show that:
\begin{itemize}
\item it is possible to construct a graphical method for
      the description of the connection representation;
\item this description is formally identical to the description
      of the loop representation.
\item the recently proposed scalar product in the loop
      representation \cite{DePietri96} coincides with the
      Ashtekar Lewandowski measure \cite{FunctINT,Baez}.
\end{itemize}
The essential bridge between the loop representation and the
connection representation is given by the so called
loop-transformation \cite{LOOP} and its inverse
\cite{Thiemann96} that are rigorously defined, up to now,
only in the case of Euclidean general relativity or, which
is the same, of a real connection. Consequently, this paper
will deal with Euclidean quantum gravity. The possibility
of using a real connection also for Lorentzian gravity 
is object of intensive investigation 
(see for example \cite{Real}). 
Moreover, Thiemann and Ashtekar \cite{TA}
have argued that the most promising strategy for implementing
the quantum reality conditions is to start from the real 
Ashtekar connection and circumvent the difficulties due to the 
complicate form of the Lorentzian Hamiltonian constraint by expressing 
it in terms of the Riemannian Hamiltonian constraint via a  
{\it generalized Wick transform}.

The structure of the paper is as follows.
In section II  the basic elements of Penrose's binor
calculus are given in a form suited for quantum gravity. In section III
the basic elements of the theory of generalized measure in 
the space of connections modulo
gauge transformations $\overline{{\cal A}/{\cal G}}$
are summarized.  In section IV the definition of 
spin-network cylindrical functions is given and 
their graphical binor representations is constructed, 
while their normalization is computed in section V.  
In section VI the graphical representation of the regularized
$\tilde{E}^a_i(x)$ operator is given. In section VII these
results are used to re-derive the full spectrum of the area operator.
Finally, in section VIII the loop transformation is discussed and 
it is shown that the scalar product proposed in \cite{DePietri96} and 
Ashtekar-Lewandowski's coincide.


\section{Penrose's graphical method for 
         $SU(2)$ tensor calculus}

As shown in reference \cite{DePietri96} the loop
representation of quantum gravity is strictly related to the
graphical methods of $SU(2)$ calculus \cite{GraphMethods}
and more precisely to Penrose's binor calculus\cite{binor}. The existence
of this relationship between the loop representation and
$SU(2)$ graphical calculus constitutes the basic technical tool 
that allows us to establish an  exact relation between the
loop-representation and the connection representation. 
In \cite{DePietri96,Rovelli95a} it was shown that the most
efficient way of dealing with $SU(2)$ ($SL(2,C)$) calculus is
the use of the binor-formalism \cite{binor} to which we
refer for more detail. Here, we only recall the essential
elements of the binor graphical calculus to fix the
notation. The main idea behind this method is to rewrite any
tensor expression in which there are sums of dummy indices in
a graphical way \cite{binor}. Penrose represents the
basic elements of spinor calculus (i.e., tensor expression
with indices $A,B,\ldots=1,2$) as
\begin{equation}
\begin{array}{rclcrclcrcl}
\delta_C^{~A} &=& 
   \begin{array}{c}\setlength{\unitlength}{1 pt}
   \begin{picture}(10,20)
            \put(5,5){\line(0,1){10}}
            \put(5,5){\circle*{3}}\put(5,15){\circle*{3}}
            \put(6,0){${\scriptstyle C}$}\put(6,14){${\scriptstyle A}$}
   \end{picture}\end{array} 
&&{\rm i} \epsilon_{AC} &=& 
   \begin{array}{c}\setlength{\unitlength}{1 pt}
   \begin{picture}(20,15)
               \put(5,5){\circle*{3}}\put(15,5){\circle*{3}}
               \put(10,5){\oval(10,20)[t]}
               \put(3,0){${\scriptstyle A}$}\put(12,0){${\scriptstyle C}$}
   \end{picture}\end{array}
&&{\rm i} \epsilon^{AC} &=& 
   \begin{array}{c}\setlength{\unitlength}{1 pt}
   \begin{picture}(20,15)
              \put(10,10){\oval(10,20)[b]}
              \put(5,10){\circle*{3}}\put(15,10){\circle*{3}}
              \put(3,11){${\scriptstyle A}$}\put(12,11){${\scriptstyle C}$}
   \end{picture}\end{array} \\
\eta_{A} &=& 
   \begin{array}{c}\setlength{\unitlength}{1 pt}
   \begin{picture}(20,15)
               \put(0,5){\framebox(10,10){$\eta$}}
               \put(6,-2){${\scriptstyle A}$}
               \put(5,0){\line(0,1){5}}\put(5,0){\circle*{3}}
   \end{picture}\end{array}
&&\eta^{A} &=& 
   \begin{array}{c}\setlength{\unitlength}{1 pt}
   \begin{picture}(20,15)
              \put(0,0){\framebox(10,10){$\eta$}}
              \put(6,13){${\scriptstyle A}$}
              \put(5,10){\line(0,1){5}}\put(5,15){\circle*{3}}
   \end{picture}\end{array}
&& X_{AB}^{C} &=& 
   \begin{array}{c}\setlength{\unitlength}{1 pt}
   \begin{picture}(20,25)
      \put(0,7){\framebox(20,10){${\scriptstyle X}$}}
      \put(6,0){${\scriptstyle A}$}\put(5,2){\line(0,1){5}}
      \put(5,2){\circle*{3}}
      \put(16,0){${\scriptstyle B}$}\put(15,2){\line(0,1){5}}
      \put(15,2){\circle*{3}}
      \put(11,20){${\scriptstyle C}$}\put(10,17){\line(0,1){5}}
      \put(10,22){\circle*{3}}
   \end{picture}\end{array}
\end{array}
\label{eq:binCONV}
\end{equation}
and assigns to any crossing\footnote{The binor representation is
a graphical representation in the plane.} a minus sign, i.e:
\begin{equation}
\delta_C^{~A}\delta_D^{~B} = - 
   \begin{array}{c}\setlength{\unitlength}{1 pt}
   \begin{picture}(20,20)
        \put(5,5){\line(1,1){10}}
        \put(5,5){\circle*{3}}\put(5,15){\circle*{3}}
        \put(6,0){${\scriptstyle C}$}\put(6,14){${\scriptstyle A}$}
        \put(15,5){\line(-1,1){10}}
        \put(15,5){\circle*{3}}\put(15,15){\circle*{3}}
        \put(16,0){${\scriptstyle D}$}\put(16,14){${\scriptstyle B}$}
   \end{picture}\end{array}
\label{eq:minus}
\end{equation}
Using this rule it is possible to represent any $SU(2)$
($SL(2,C)$) tensor expression in a graphical way as follows:
({\bf 1}) define the  ``up'' direction in the plane;
({\bf 2}) draw boxes with the name of the $SU(2)$ tensor
   (except for the $\delta$ and the $\epsilon$ that are simply
   represented by lines) with as many slots going up as the
   number of contravariant indices and as many slots going down as
   the number of covariant indices; 
({\bf 3}) for each dummy index connect with a line the
          corresponding slots of the boxes;
({\bf 4}) assign a ``${\rm i}$'' factor to each minimum or maximum
          of the lines;
({\bf 5}) assign a minus sign to each crossings of lines.  

Conversely, any curve can be decomposed in a product of $\delta$'s and
$\epsilon$'s. It is possible to prove that two curves
that are ambient isotopic, i.e., that can be transformed one
into the other by a sequence of Reidemeister
\cite{Reidemeister} moves, represent the same tensorial
expression as a product of epsilons and deltas, and, indeed, two
drawings correspond to the same tensor expression if they can
be transformed one into the other by a sequence of Reidemeister
moves and translations (not rotations) of the boxes 
representing the true tensors.

A closed loop (with this convention) has value $(-2)$, 
($\begin{array}{c}\setlength{\unitlength}{1 pt}\begin{picture}(10,10)
  {\put(5,5){\oval(10,10)}}\end{picture}\end{array}
  = {\rm i} \epsilon_{AB}\ {\rm i} \epsilon^{AB} = - 2
$) and the identity $\epsilon_{AC} \epsilon^{BC} = \delta_A^C$
reads
\begin{equation}
    \begin{array}{c}\setlength{\unitlength}{1 pt}
    \begin{picture}(10,15)
        \put( 5, 0){\line(0,1){15}}
    \end{picture}\end{array}
= - \begin{array}{c}\setlength{\unitlength}{1 pt}
    \begin{picture}(20,20)
        \put(10,10){\oval(10,10)[t]}
        \put(10, 5){\oval(10,10)[b]}
        \put( 5, 5){\line(0,1){5}}
     \end{picture}\end{array}
~~.
\end{equation}
A trace of a matrix will be written as
\begin{equation}
 {\rm Tr} X^A_{~B} = \delta^B_A X^A_{~B} 
 = - \begin{array}{c}\setlength{\unitlength}{1 pt}
     \begin{picture}(20,20)
        \put(10,15){\oval(10,10)[t]}
        \put(10, 5){\oval(10,10)[b]}
        \put(10, 5){\framebox(10,10){$\scriptstyle X$}}
        \put( 5, 5){\line(0,1){10}}
     \end{picture}\end{array}
\label{eq:binorTRACE}
~~,
\end{equation}
and indeed a closed loop denotes the operation of taking
$(-1)$ the trace of the corresponding 
tensor expression.
Moreover, the basic binor identity is graphically given by
\begin{equation}
\fl \quad
   \begin{array}{c}\setlength{\unitlength}{1 pt}
   \begin{picture}(10,15)
            \put( 0,0){\line( 2,3){10}}
            \put(10,0){\line(-2,3){10}}
   \end{picture}\end{array} 
+  \begin{array}{c}\setlength{\unitlength}{1 pt}
   \begin{picture}(10,15)
            \put( 0,0){\line(0,1){15}}
            \put(10,0){\line(0,1){15}}  
   \end{picture}\end{array}
+  \begin{array}{c}\setlength{\unitlength}{1 pt}
   \begin{picture}(10,15)
            \put(5, 0){\oval(10,10)[t]}
            \put(5,15){\oval(10,10)[b]} 
   \end{picture}\end{array}
= (-1)   ~\delta^{\cdot C}_{B}\ \delta^{\cdot D}_{A}  
   +      ~\delta^{\cdot C}_{A}\ \delta^{\cdot D}_{B}
   + (-1) ~\epsilon_{AB}\ \epsilon^{CD} = 0  
\nonumber
\end{equation}
Clearly, in Penrose's binor calculus there is no meaning in the
distinction between over and under crossing. In Penrose's binor
notation it is also possible to write any expression
involving representations of the $SU(2)$ group.  In particular,
the irreducible representation $\pi_i(n_i)$ (labeled by an
integer $n$, its color, that is twice the spin: $n = 2 j_n$) can be
constructed as the symmetrization $\Pi^{(e)}_n$ (in Penrose's
graphical representation an anti-symmetrization because of equation
(\ref{eq:minus})) of the
tensor product of $n$ fundamental representation
$\begin{array}{c}\setlength{\unitlength}{1 pt}
 \begin{picture}(6,10)
    \put(3,0){\line(0,1){2}}\put(3,8){\line(0,1){2}}
    \put(0,2){\framebox(6,6){$\scriptscriptstyle g$}}
 \end{picture}\end{array}
{}_A^B = U(g)_A^B$ ($g$ being the group element).
Denoting by $\Pi^{(e)}_n$ the normalized symmetrizer
(graphically an anti-symmetrizer), the color $n$ irreducible representation
is given by:
\begin{eqnarray}
&&\Pi^{(e)}_n P_\alpha = \frac{1}{n!} \sum_p
(-1)^{|p|}\ P^{(p)}_{\alpha}
=  \begin{array}{c}\setlength{\unitlength}{1 pt}
   \begin{picture}(20,25)
     \put(10, 0){\line(0,1){10}}
     \put(0,10){\framebox(20,5){}}
     \put(10,15){\line(0,1){10}}\put(12,17){n}
   \end{picture}\end{array}
\\
&& \pi_i(n_i)   
=  \begin{array}{c}\setlength{\unitlength}{1 pt}
   \begin{picture}(15,30)
    \put(7,22){$\scriptstyle n_i$}
    \put(10,12){$\scriptstyle {e_i}$}
    \put(2,5){\rule{6pt}{15pt}}
    \put(5,0){\line(0,1){5}}
    \put(5,20){\line(0,1){5}}
   \end{picture}\end{array}
=  \begin{array}{c}\setlength{\unitlength}{1 pt}
   \begin{picture}(40,40)
    \put(22,2){$\scriptstyle n_i$}
    \put(0,10){\framebox(40,2){}} \put(20,0){\line(0,1){10}}
    \put(5,12){\line(0,1){3}}\put(5,25){\line(0,1){3}}
    \put(-1,15){\framebox(12,10){$\scriptstyle g_{e_i}$}}
    \put(12,18){$\cdots$}
    \put(35,12){\line(0,1){3}}\put(35,25){\line(0,1){3}}
    \put(29,15){\framebox(12,10){$\scriptstyle g_{e_i}$}}
    \put(0,28){\framebox(40,2){}} \put(20,30){\line(0,1){10}}
    \put(22,34){$\scriptstyle n_i$}
   \end{picture}\end{array}
~, 
\label{eq:5}
\end{eqnarray}
($|p|$ is the parity of the permutation and a line labeled by a
positive integer $n$ represents $n$ non-intersecting parallel
line). The only additional needed informations about this formalism
are: ({\it i})  an explicit graphical representation of the Clebsh-Gordon
intertwining matrix i.e., of the matrix that represents the coupling of $3$
(or more) irreducible representations of the group $SU(2)$, and
({\it ii}) the concept of chromatic evaluation\cite{Cromatic}
(all the other properties can be deduced from these).  
The representation of the Clebsh-Gordon intertwining
matrix in the binor formalism is
given by the special sum of ``tangles'' denoted as the
$3$-vertex. Each line of the vertex is labeled by a
positive integer $a$, $b$ or $c$ and is defined as:
\begin{equation}
   \begin{array}{c}\setlength{\unitlength}{1 pt}
   \begin{picture}(30,40)
       \put(15,15){\line(-1, 1){10}} \put( 4,27){$a$}
       \put(15,15){\line( 1, 1){10}} \put(22,27){$b$}
       \put(15, 5){\line(0,1){10}}   \put(17,1){$c$}
       \put(15,15){\circle*{3}}
   \end{picture}\end{array}
\stackrel{def}{=}
   \begin{array}{c}\setlength{\unitlength}{1 pt}
   \begin{picture}(70,40) 
       \put(20,30){\line( 1, 0){30}} \put(28,32){$m$}
       \put(35,15){\line(-1, 1){15}} \put(20,17){$p$}
       \put(35,15){\line( 1, 1){15}} \put(44,17){$n$}
       \put(35, 1){\line(0,1){10}}   \put(42,1){$c$}
       \put(27,11){\framebox(16,4){}}
       \put(16,22){\framebox(4,16){}} 
       \put(6 ,30){\line(1,0){10}}\put(0,30){a}
       \put(50,22){\framebox(4,16){}}
       \put(54,30){\line(1,0){10}}\put(65,30){b}
   \end{picture}\end{array}
\;\;
\left\{\begin{array}{rcl} 
m &=& (a+b-c)/2 \\ n&=&(b+c-a)/2 \\ p&=&(c+a-b)/2
\end{array}\right.
\label{eq:6}
\end{equation}
where it is assumed that $m,n,p$ are positive integers.  This
last condition is called the {\it admissibility condition}
for the $3$-vertex $V_3(a,b,c)$.  Moreover, by the Wigner-Eckart
theorem, or more precisely the version of it due to Yutsin,
Levinson and Vanagas \cite{GraphMethods}, any invariant tensorial
intertwining matrix representing the coupling of $n$
representations of a compact group can be expressed as the
product of Clebsh-Gordon coefficients or, which is the same,
in terms of a three-valent decomposition. It is easy to see
that any contractor $c_i$ of a given vertex $v_i$ with $n$
incoming (outgoing) edges, can be written in terms of a
trivalent decomposition labeled by ($n$-$3$) integers; graphically:
\begin{equation}
\begin{array}{c}\setlength{\unitlength}{1 pt}
\begin{picture}(100,55)
     \put( 0, 0){$P_0$}\put(15, 0){\line(2,1){20}}
     \put( 0,20){$P_1$}\put(15,20){\line(1,0){20}}
     \put(10,40){$P_2$}\put(25,40){\line(1,-1){10}}
     \put(35,0){\framebox(25,30){$c_i$}}
     \put(40,40){$\cdots$}        \put(40,33){$\ldots$} 
     \put(75,40){$P_{n-3}$}\put(60,30){\line(1,1){10}}
     \put(75,20){$P_{n-2}$}\put(60,20){\line(1,0){10}}
     \put(75, 0){$P_{n-1}$}\put(60,10){\line(1,-1){10}} 
   \end{picture}\end{array}
=  \begin{array}{c}\setlength{\unitlength}{1 pt}
   \begin{picture}(110,55)
     \put( 0, 0){$P_0$}\put(15, 0){\line(1,1){10}}
     \put( 0,20){$P_1$}\put(15,20){\line(1,0){10}}
     \put(10,40){$P_2$}\put(25,40){\line(1,-1){10}}
     \put(25,10){\line(0,1){10}}
     \put(25,20){\line(1,1){10}}\put(32,18){$\scriptstyle i_2$}
     \put(35,30){\line(1,0){10}}\put(40,22){$\scriptstyle i_3$}
     \put(25,20){\circle*{3}}   \put(35,30){\circle*{3}}
     \put(45,40){$\cdots$}       \put(45,30){$\ldots$} 
     \put(60,30){\line(1, 0){10}}
     \put(70,30){\line(1,-1){10}}
     \put(60,20){$\scriptstyle i_{n-2}$}
     \put(80,20){\line(0,-1){10}} 
     \put(80,20){\circle*{3}}\put(70,30){\circle*{3}}
     \put(85,40){$P_{n-3}$}  \put(70,30){\line(1,1){10}}
     \put(95,20){$P_{n-2}$} \put(80,20){\line(1,0){10}}
     \put(95, 0){$P_{n-1}$} \put(80,10){\line(1,-1){10}} 
   \end{picture}\end{array}
~.
\label{eq:7}
\end{equation}
It is important to note that the orientation of the colored 
line connected to a three vertex has to be specified since
\begin{equation}
\fl
\quad
   \begin{array}{c}
   \setlength{\unitlength}{1 pt}
   \begin{picture}(40,50)
       \put(30,25){\line(-3, 2){30}} \put( 0,35){$a$}
       \put( 0,25){\line( 3, 2){13}} 
       \put(30,45){\line(-3,-2){13}} \put(30,35){$b$}
       \put(15,25){\oval(30,20)[b]}  \put(15,15){\circle*{3}}
       \put(15, 5){\line(0,1){10}}   \put(17,1){$c$}
   \end{picture}\end{array} 
      =\lambda^{ab}_c 
   \begin{array}{c}\setlength{\unitlength}{1 pt}
   \begin{picture}(40,40)
       \put(15,15){\line(-1, 1){10}} \put( 4,27){$a$}
       \put(15,15){\line( 1, 1){10}} \put(22,27){$b$}
       \put(15, 5){\line(0,1){10}}   \put(17,1){$c$}
       \put(15,15){\circle*{3}}
    \end{picture}\end{array}
\quad\mbox{where}\qquad  
\lambda^{ab}_c = (-1)^{\textstyle
     \frac{a(a+3)+b(b+3)-c(c+3)}{2}}
~~.
\end{equation}

The Wigner-Eckart theorem allows to exchange, in a contractor
$c_i$, the order in which the colored lines are connected by 
three-vertices. This possibility is
expressed by the recoupling theorem 
\cite{GraphMethods,Cromatic,Citanovic} that states
\begin{equation}
\begin{array}{c}\setlength{\unitlength}{1 pt}
\begin{picture}(50,40)
          \put( 0,0){$a$}\put( 0,30){$b$}
          \put(45,0){$d$}\put(45,30){$c$}
          \put(10,10){\line(1,1){10}}\put(10,30){\line(1,-1){10}}
          \put(30,20){\line(1,1){10}}\put(30,20){\line(1,-1){10}}
          \put(20,20){\line(1,0){10}}\put(22,25){$j$}
          \put(20,20){\circle*{3}}\put(30,20){\circle*{3}}
\end{picture}\end{array}
    = \sum_i  \left\{\begin{array}{ccc}
                      a  & b & i \\
                      c  & d & j  
              \end{array}\right\}
\begin{array}{c}\setlength{\unitlength}{1 pt}
\begin{picture}(40,40)
      \put( 0,0){$a$}\put( 0,40){$b$}
      \put(35,0){$d$}\put(35,40){$c$}
      \put(10,10){\line(1,1){10}}\put(10,40){\line(1,-1){10}}
      \put(20,30){\line(1,1){10}}\put(20,20){\line(1,-1){10}}
      \put(20,20){\line(0,1){10}}\put(22,22){$i$}
      \put(20,20){\circle*{3}}\put(20,30){\circle*{3}}
\end{picture}\end{array}
\label{eq:REC}
\end{equation}
where the quantities 
$\left\{\begin{array}{ccc} a & b & i \\  c & d & j
\end{array}\right\}$ are the Wigner $SU(2)$ six-j symbols.

We are now ready to express the concept of chromatic
evaluation.  Consider two lines of color $n$. They represent
a tensor expression with $n$-indices, and indeed they can be
connected to one another to represent the sum of $n$
dummy indices. Clearly, two lines can be connected only if they
have the same color. Let me consider two vertices of type
(\ref{eq:7}) $V_n(a_1,\ldots,a_n)$ and
$V_n^\prime(b_1,\ldots,b_n)$ with $n$-connected lines.  If all
the corresponding lines have the same color
($a_i=b_i$ for $i=1,\ldots,n$), it is possible to attach 
the $a_i$ lines and the corresponding $b_i$ lines
to one another. In this way
we obtain a closed scalar expression of $\delta$
and $\epsilon$ tensors in which all the dummy indices
are summed, i.e., a number. The
evaluation of this number is, by definition, the chromatic
evaluation of a network. The chromatic evaluation of a network
obtained by joining two vertices will be denoted by
${\langle}V_n(a_1,\ldots,a_n)|V_n^\prime(b_1,\ldots,b_n)\rangle$
and it is a scalar product in the space
of the possible contractors $c_i$ of a vertex.  In particular,
the chromatic evaluation of a $2$-vertex (i.e. of a line) 
and that of a $3$-vertex will be defined as
$\Delta$ function  and $\theta$ function, respectively:
\begin{eqnarray}
  \Delta_n &=& \langle V_2(n),V_2(n) \rangle \! 
  =\!\begin{array}{c}\setlength{\unitlength}{.5 pt}
     \begin{picture}(35,25)
        \put(15, 0){\line(0,1){10}}\put(20, 0){\line(0,1){10}}
        \put(15, 0){\line(1,0){ 5}}\put(15,10){\line(1,0){ 5}}
        \put(15,25){\line(1,0){5}}  \put(15,15){$\scriptstyle n$}
        \put(15,15){\oval(30,20)[l]}\put(20,15){\oval(30,20)[r]} 
     \end{picture}\end{array} \!
= (-1)^n (n+1)
\\
\theta(a,b,c) 
  &=&  \langle V_3(a,b,c),V_3(a,b,c) \rangle 
   = \begin{array}{c}\setlength{\unitlength}{.5 pt}
     \begin{picture}(40,40)
        \put(18,32){$\scriptstyle a$}
        \put( 0,15){\line(1,0){40}} \put(18,17){$\scriptstyle b$}
        \put(20,15){\oval(40,30)}   \put(18, 2){$\scriptstyle c$}
        \put( 0,15){\circle*{3}}    \put(40,15){\circle*{3}}
     \end{picture}\end{array}
~~~~.
\end{eqnarray} 
Using this formalism, the anti-hermitian generator
of the $SU(2)$ group $\tau_i{}^A_{~B} = \frac{{\rm i}}{2}
\sigma_i{}^A_{~B} $ ($\sigma_i$, Pauli matrix) will be denoted by
\begin{equation}
\tau_i{}^A_{~B}
=  \begin{array}{c}\setlength{\unitlength}{1 pt}
   \begin{picture}(8,18)
    \put(4,0){\line(0,1){2}}\put(4,10){\line(0,1){2}}
    \put(4,6){\circle{8}} \put(2,4){$\scriptstyle i$}
   \end{picture}\end{array}{}^A_{~B}
~~.
\label{eq:Deftau}
\end{equation}
From equation (\ref{eq:Deftau}) we obtain the identity
\begin{eqnarray}
  \sum_{i=1}^3 \tau_i{}^A_{~B}  \tau_i{}^C_{~D}  
&=& \sum_{i=1}^3  
   \begin{array}{c}\setlength{\unitlength}{1 pt}
   \begin{picture}(20,20)
    \put(4,0){\line(0,1){5}}\put(4,13){\line(0,1){4}}
    \put(4,9){\circle{8}} \put(2,7){$\scriptstyle i$}
    \put(5,0){$\scriptstyle B$} \put(5,14){$\scriptstyle A$}
    \put(14,0){\line(0,1){5}}\put(14,13){\line(0,1){4}}
    \put(14,9){\circle{8}} \put(12,7){$\scriptstyle i$}
    \put(15,0){$\scriptstyle D$} \put(15,14){$\scriptstyle C$}
  \end{picture}\end{array}
= - \frac{1}{4} \bigg[
- \begin{array}{c}\setlength{\unitlength}{1 pt}
  \begin{picture}(10,25)
          \put( 0,0){\line( 2,3){10}}
          \put(10,0){\line(-2,3){10}} 
  \end{picture}\end{array}
+ \begin{array}{c}\setlength{\unitlength}{1 pt}
  \begin{picture}(10,25)
           \put(5, 0){\oval(10,10)[t]}
           \put(5,15){\oval(10,10)[b]} 
   \end{picture}\end{array}
 \bigg]
\nonumber\\
&=& - \frac{1}{2} 
   \begin{array}{c}\setlength{\unitlength}{1 pt}
   \begin{picture}(30,30)
     \put( 0, 0){\line(1,1){10}} \put( 0,22){\line(1,-1){10}}
     \put(10,10){\line(1,0){2}}  \put(17,10){\line(1,0){3}} 
     \put(10,12){\line(1,0){2}}  \put(17,12){\line(1,0){3}} 
     \put(12, 0){\framebox(5,22){}} \put(12,24){$\scriptstyle 2$}
     \put(30, 0){\line(-1,1){10}}\put(30,22){\line(-1,-1){10}}
   \end{picture}\end{array}
= - \frac{1}{2} 
   \begin{array}{c}\setlength{\unitlength}{1 pt}
   \begin{picture}(30,30)
     \put( 0, 5){\line(1,1){10}} \put( 0,25){\line(1,-1){10}}
     \put(10,15){\line(1,0){10}} 
     \put(14,17){$\scriptstyle 2$}
     \put(30, 5){\line(-1,1){10}}\put(30,25){\line(-1,-1){10}}
     \put(10,15){\circle*{3}}     \put(20,15){\circle*{3}}
     \put( 0,25){$\scriptstyle A$}\put(22,25){$\scriptstyle C$}
     \put( 0, 0){$\scriptstyle B$}\put(22, 0){$\scriptstyle D$}
   \end{picture}\end{array}  
\end{eqnarray}
which can be written as
\begin{equation}
\sum_{i=1}^3 \;\;
\begin{array}{c}\setlength{\unitlength}{1 pt}
\begin{picture}(40,60)
    \put( 2,12){$\scriptstyle 2$} 
    \put( 0,10){\line(1,0){10}} \put(10,10){\circle*{3}}
    \put(10, 5){\line(0,1){10}}
    \put(15, 5){\oval(10,10)[b]} \put(15,15){\oval(10,10)[t]}
    \put(20,10){\circle{10}}     \put(19, 7){$\scriptstyle i$}
    \put( 2,42){$\scriptstyle 2$} 
    \put( 0,40){\line(1,0){10}}   \put(10,40){\circle*{3}}
    \put(10,35){\line(0,1){10}}
    \put(15,35){\oval(10,10)[b]}  \put(15,45){\oval(10,10)[t]}
    \put(20,40){\circle{10}}      \put(19,37){$\scriptstyle i$}
\end{picture}\end{array}
= \frac{1}{2}\;\;
\begin{array}{c}\setlength{\unitlength}{1 pt}
\begin{picture}(30,60)
    \put( 2,12){$\scriptstyle 2$} 
    \put( 0,10){\line(1,0){10}} \put(10,10){\circle*{3}}
    \put( 2,42){$\scriptstyle 2$} 
    \put( 0,40){\line(1,0){10}} \put(10,40){\circle*{3}}
    \put(10,25){\oval(30,30)[r]}
\end{picture}\end{array}
~~~.
\label{eq:2tau}
\end{equation}
Analogously, the following relation is also true
\begin{equation}
\sum_{i,j,k=1}^3 \;\epsilon_{ijk}\;
\begin{array}{c}\setlength{\unitlength}{1 pt}
\begin{picture}(40,60)
    \put( 2,12){$\scriptstyle 2$} 
    \put( 0,10){\line(1,0){10}} \put(10,10){\circle*{3}}
    \put(10, 5){\line(0,1){10}}
    \put(15, 5){\oval(10, 5)[b]} \put(15,15){\oval(10, 5)[t]}
    \put(20,10){\circle{10}}     \put(19, 7){$\scriptstyle i$}
    \put( 2,32){$\scriptstyle 2$} 
    \put( 0,30){\line(1,0){10}}   \put(10,30){\circle*{3}}
    \put(10,25){\line(0,1){10}}
    \put(15,25){\oval(10, 5)[b]}  \put(15,35){\oval(10, 5)[t]}
    \put(20,30){\circle{10}}      \put(19,27){$\scriptstyle j$}
    \put( 2,52){$\scriptstyle 2$} 
    \put( 0,50){\line(1,0){10}}   \put(10,50){\circle*{3}}
    \put(10,45){\line(0,1){10}}
    \put(15,45){\oval(10, 5)[b]}  \put(15,55){\oval(10, 5)[t]}
    \put(20,50){\circle{10}}      \put(19,47){$\scriptstyle k$}
\end{picture}\end{array}
= \frac{1}{2}\;\;
\begin{array}{c}\setlength{\unitlength}{1 pt}
\begin{picture}(40,60)
    \put( 2,12){$\scriptstyle 2$} 
    \put( 0,10){\line(1,0){10}} \put(10,10){\circle*{3}}
    \put( 2,32){$\scriptstyle 2$} 
    \put( 0,30){\line(1,0){10}}   \put(10,30){\circle*{3}}
    \put( 2,52){$\scriptstyle 2$} 
    \put( 0,50){\line(1,0){10}}   \put(10,50){\circle*{3}}
    \put(10,30){\oval(20,40)[r]}
    \put(10,30){\line(1,0){10}}
    \put(20,30){\circle*{3}}
\end{picture}\end{array}
~~~.
\label{eq:3tau}
\end{equation}
The Penrose's binor formalism allows to graphically 
write the fundamental property 
\begin{equation}
U^{-1}{}^A_{~B} =  \epsilon^{AC} \epsilon_{BD} U^D_{~C}
~~,
\end{equation}
valid for any $2\times 2$ matrix
with unit determinant, as
\begin{equation}
\begin{array}{c}\setlength{\unitlength}{1 pt}
\begin{picture}(15,30)
    \put(7,0){\line(0,1){8}}\put(7,22){\line(0,1){8}}
    \put(0,8){\framebox(14,14){$\scriptstyle X^{\scriptscriptstyle -1}$}}
    \put(9,0){$\scriptstyle B$}\put(9,24){$\scriptstyle A$}
\end{picture}\end{array}
= 
\begin{array}{c}\setlength{\unitlength}{1 pt}
\begin{picture}(40,30)
    \put(5,0){\line(0,1){22}}\put(7,0){$\scriptstyle B$}
    \put(12,22){\oval(14,16)[t]}
    \put(12,8){\framebox(14,14){$\scriptstyle X$}}
    \put(26,8){\oval(14,16)[b]}
    \put(33,8){\line(0,1){22}} \put(35,24){$\scriptstyle A$}
\end{picture}\end{array}
\;\; {\rm and} \;\;
\begin{array}{c}\setlength{\unitlength}{1 pt}
\begin{picture}(25,30)
    \put(7,22){$\scriptstyle n_i$}
    \put(10,12){${e_i}^{\scriptscriptstyle -1}$}
    \put(2,5){\rule{6pt}{15pt}}
    \put(5,0){\line(0,1){5}}
    \put(5,20){\line(0,1){5}}
\end{picture}\end{array} 
=  
\begin{array}{c}\setlength{\unitlength}{1 pt}
\begin{picture}(45,30)
    \put(37,2){$\scriptstyle n_i$}
    \put(5,5) {\line(0,1){20}}
    \put(20,12){${e_i}$}
    \put(12,5){\rule{6pt}{15pt}}
    \put(10, 5){\oval(10,10)[b]}
    \put(25,20){\oval(20,10)[t]}
    \put(35,0){\line(0,1){20}}
\end{picture}\end{array}
~~. 
\label{eq:edgerevers}
\end{equation}
These identities will play an essential role in the following.
For any other detail about binor calculus, its relation
to the theory of angular momentum and the explicit value
of the symbols, we refer to \cite{Kauffman94}
and to the appendices of \cite{DePietri96}.


\section{Connection representation and integration on
the space of connections modulo gauge transformations}

In this section we give a synthetic review of the
construction used in
\cite{Ashtekar95a,FunctINT,Marolf95,Ashtekar94}.
This construction can be summarized as follows: 
({\bf i}) the space of histories for
   quantum gauge field theory $\overline{{\cal A}/{\cal G}}$ is
   taken to be the Gel'fand spectrum generated by the Wilson
   loop functionals; 
({\bf ii}) in this space a measure $d\mu_0(A)$ is naturally defined 
as the $\sigma$-additive extension of the family of measures 
$d\mu_{0,\gamma}(A)=d\mu_H(g_{e_1})\ldots d\mu_H(g_{e_n})$
in the space $\overline{{\cal A}/{\cal G}}_\gamma$
\cite{Marolf95} of the cylindrical
functions associated to piecewise analytical
graphs $\gamma$; ({\bf iii}) a natural basis in 
the space $\overline{{\cal A}/{\cal G}}$
is given by the spin-network states \cite{Baez}.
The relationship of this procedure to the standard method of
constructive field theory is discussed in
\cite{Ashtekar94b}. In more detail, a 
function $f_\gamma$ ($f_\gamma \in \overline{{\cal A}/{\cal G}}_\gamma$)
is said to be cylindrical with respect to a graph\footnote{ $\gamma$
is assumed to be an analytic, i.e. each of its edges is
an analytic embedding of $M^3$.} 
$\gamma$ if it is a gauge
invariant function of the finite set of arguments
$(g_{e_1}(A),\ldots,g_{e_n}(A))$ where the $g_{e_i} =
{\cal P}\exp(-\int_{e_i} A)$ are the holonomies of $A$ along the edges
$e_i$ of the graph $\gamma$. Given these families (one for each
analytic graph $\gamma$ embedded in $\Sigma$) of integrable
functions (the cylindrical ones), the measure on
these finite dimensional projections 
$\overline{{\cal A}/{\cal G}_\gamma}$ 
(which are isomorphic to $G^n$) is given by
\begin{eqnarray}
&&\int_{\overline{{\cal A}/{\cal G}}} d\mu_0(A) f_\gamma(A) 
  = \int_{\overline{{\cal A}/{\cal G}}_\gamma} 
          d\mu_{0,\gamma}(A) f_\gamma(A)
  =
\label{eq:defINT}
\\
&&\qquad =\int_{G^n} d\mu_H(g_{e_1})\ldots d\mu_H(g_{e_n})
           ~ f_\gamma(g_{e_1},\ldots,g_{e_n})
~~.
\nonumber
\end{eqnarray} 
Since this family of finite dimensional measures verifies
the consistency condition of \cite{NEW}, it has
a $\sigma$-additive extension $d\mu_0(A)$ that 
defines a generalized measure on the whole space $\overline{{\cal
A}/{\cal G}}$.  The quantum configuration space can be
assumed to be the space $L^2[{\overline{{\cal A}/{\cal G}}},d\mu_0(A)]$
of the square integrable functions with respect to this
measure.
 

\section{Spin-networks and their graphical representation}

Following Baez\footnote{Precisely, we are using
Thiemann's\cite{Thiemann96} definition of edge-network. 
We point out that there
is no privileged or fundamental role in the $3$-valent
decomposition; it is only a useful way of
constructing a basis in the space of all the inequivalent
contractors that can be assigned to a vertex.}  \cite{Baez}, 
given a triple $S=(\gamma,\vec{\pi},\vec{c})$,
a {\it spin-network} state is the cylindrical map (with
respect to the graph $\gamma$) from $\overline{{\cal A}/{\cal
G}}_\gamma$ into $C$ defined by:
\begin{equation}
{\cal T}_{\gamma,\vec{\pi},\vec{c}}[A] \stackrel{def}{=}
  \left( \otimes_{i=1}^N \pi_i(g_{e_i}) \right) \cdot
  \left( \otimes_{j=1}^M c_j \right) 
~~,
\label{eq:defSPINnet}
\end{equation}
where the elements of the triple $S=(\gamma,\vec{\pi},\vec{c})$ are:
\begin{description} 
\item[{\bf (i)}] a graph $\gamma$ with 
   $N$ edges ${\cal E}= \{ e_1,\ldots,e_N\}$ and 
   $M$ vertices ${\cal V} = \{ v_1,\ldots,v_M \}$; 
\item[{\bf (ii)}] a labeling $\vec{\pi}=(\pi_1,\ldots,\pi_N)$ of 
   the edges $e_1,\ldots,e_N\in{\cal E}$ of $\gamma$ with 
   irreducible representation $\pi_i$ of $G$; 
\item[{\bf (iii)}] a labeling $\vec{c}=(c_1,\ldots,c_M)$ of
    the vertices $v_1,\ldots,v_M\in {\cal V}$ of $\gamma$ 
    with contractors $c_j$ (the intertwining matrices $c_j$, 
    in each of the  vertices $v_j$, represent the coupling of the $n_j$ 
representations associated to the $n_j$ edges that 
start or end in $v_j$),
\end{description}

Since the group involved is $SU(2)$, any irreducible
unitary representation is labeled by an integer $n$ (its
color $n$ that is twice the spin: $n = 2 j_n$) and is
given by the completely symmetric tensor product of $n$
irreducible color $1$ (spin $1/2$) fundamental
representations (see equation \ref{eq:5}). Moreover, from the
discussion of section II (equations (\ref{eq:6}) and
(\ref{eq:7})), it follows that any contractor $c_j$ of the
$n_j$ valent vertex $v_j$ is uniquely determined by an
ordering of the incoming (outgoing) edges and by 
the $n_j$-$3$ integers labeling its trivalent decomposition. 
To complete the definition, the normalization choice for the
irreducible representations $\pi_i(e_i)$ and for the
contractors $c_i$ are given by equations
(\ref{eq:5}),(\ref{eq:6}) and (\ref{eq:7}), respectively.  

\begin{figure}
\begin{center}
\mbox{\psfig{file=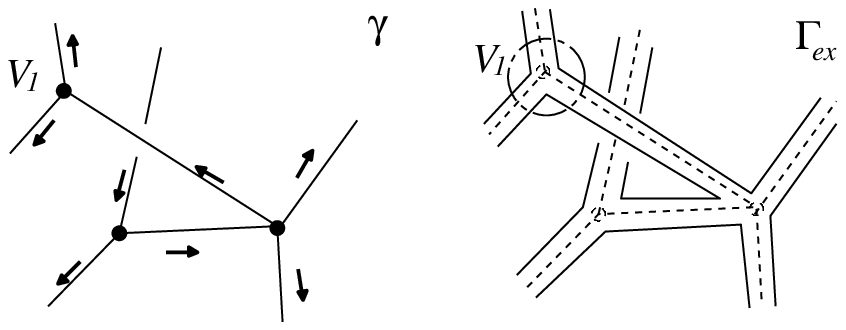}} 
\end{center}
\caption{The graph $\gamma$ and a possible extended-planar 
projection $\Gamma_{ex}$.}
\label{fig:graph}
\end{figure}

  This choice of the contractors has the property 
of being given in terms of a linear combinations of products of 
the real tensor $\epsilon$ and $\delta$ tensors. 
From the reality of the contractor $c_j$,  the unitarity of the 
group $SU(2)$ and equation (\ref{eq:edgerevers}) the reality of 
the spin-network cylindrical functions follows,
\begin{equation}
  {\overline{{\cal T}_{\gamma,\vec{\pi},\vec{c}}[A]}} ={\cal
   T}_{\gamma,\vec{\pi},\vec{c}}[A] ~~.
\label{eq:real}
\end{equation}
A spin-network is indeed fixed by a graph, a labeling
with positive integers of its edges and by trivalent decompositions
of the vertices
(note that these are exactly the same elements that characterize 
the spin-network states in the loop representation
\cite{Rovelli95,DePietri96}).
The graphical representation of 
a {\it spin-network} state is defined following the 
construction of the loop representation \cite{DePietri96}. 
As a first step, we consider a
projection of the graph $\gamma$ in a plane and its
extended planar graph $\Gamma_{ex}$ (see figure \ref{fig:graph}).

\begin{figure}
\begin{center}
\mbox{\psfig{file=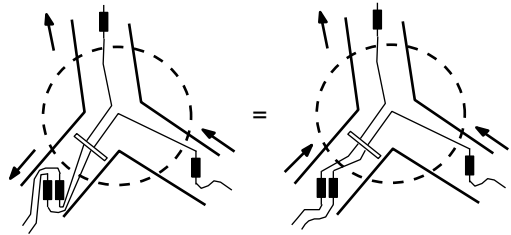}} 
\end{center}
\caption{The binor representation of a cylindrical function at the
vertex $V_1$ of graph $\gamma$ of Fig.\ 1 on the extended-planar 
projection $\Gamma_{ex}$, and its independence of the orientations
of the edges.}
\label{fig:graphbinorrep}
\end{figure}

Then, in each of its edges $e_i$, 
using equation (\ref{eq:5}), the graphical representation of 
$\pi_i(g_{e_i})$ is inserted, and in each of the vertices $v_j$,
using equations\ (\ref{eq:6}) and (\ref{eq:7}), the 
graphical representation of the corresponding contractors $c_j$
is inserted.
From equation (\ref{eq:edgerevers}) it follows that this
planar graphical representation is independent of the orientations
of the edges (see figure \ref{fig:graphbinorrep}).

\begin{figure}
\begin{center}
\mbox{\psfig{file=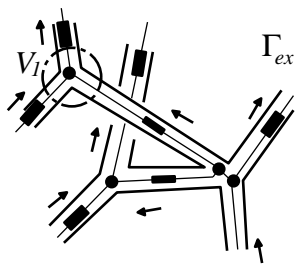}} 
\end{center}
\caption{The binor representation of a cylindrical function of
the graph $\gamma$ of Fig.~1.}
\label{fig:graph-rep}
\end{figure}

At this point the graphical, planar, binor
representation of a spin-network state is obtained. 
It is immediate to
note that, were it not for the presence of filled-boxes 
in the edges of $\gamma$, the resulting graphical 
representation (see figure \ref{fig:graph-rep}), 
would be formally identical to the
loop representation of a spin-network state
\cite{DePietri96}. 


We finally remark that the notation
${\cal T}_{\gamma,\vec{\pi},\vec{c}}[A]$ is used  instead of
the notation ${T}_{\gamma,\vec{\pi},\vec{c}}[A]$, which is
usually adopted in the works on the connection 
representation, because, in the binor representation, 
the conventions (\ref{eq:binCONV}) have the effect 
of transforming the trace in $(-1)$ times the trace of 
the corresponding tensor expression 
(see equation (\ref{eq:binorTRACE})).
This means that we have used the $C^*$ algebra generated by
$(-1)$ times the trace of the Wilson loop.
 

\section{The normalization of the spin-network states}

It is now possible to prove in 
$L^2[{\overline{{\cal A}/{\cal G}}},d\mu_0(A)]$
the orthogonality of the 
spin-network states defined in the previous section. 
Given two spin-network states 
$s= {\cal T}_{\gamma,\vec{\pi},\vec{c}}[A]$ and
$s'= {\cal T}_{\gamma',\vec{\pi}',\vec{c}'}[A]$,
there is a larger graph $\overline{\gamma}$ such that 
$\gamma  \subset \overline{\gamma}$ and 
$\gamma' \subset \overline{\gamma}$ 
(this is true only for graphs that are analytically embedded,
for a discussion of the smooth case see \cite{Smooth}).
Since both $s$ and $s'$ are cylindrical functions on 
$\overline{{\cal A}/{\cal G}}_{\overline{\gamma}}$~,
the scalar product of these states is given by 
(see equation \ref{eq:defINT}):
\begin{equation}
\fl
\qquad
  \langle{s},{s'}\rangle 
= \int d\mu_{0,\overline{\gamma}}({\cal A})
  \overline{{\cal T}_{\gamma,\vec{\pi},\vec{c}}[A]}
            {\cal T}_{\gamma',\vec{\pi}',\vec{c}'}[A] 
= \int d\mu_{0,\overline{\gamma}}({\cal A})
            {\cal T}_{\gamma,\vec{\pi},\vec{c}}[A]
            {\cal T}_{\gamma',\vec{\pi}',\vec{c}'}[A]  
~~~,
\end{equation}
where we have used equation (\ref{eq:real}).
From the definition of $d\mu_{0,\overline{\gamma}}(A)$ one
has  to integrate over the group elements associated to each 
edge. This task could be easily performed using the recoupling
theorem (\ref{eq:REC}) and the intermediate result
(see Creutz \cite{Creutz78}):
\begin{equation}
\int {d_H(g_e)}
   \Pi_{n}^{(e)}{}_{B_1\cdots B_{n}}^{\bar{B}_1\cdots\bar{B}_{n}} 
    \left[ 
    U_{A_1}^{~B_1}(g_e) \cdots U_{A_{n}}^{~B_{n}}(g_e) \right]
= \delta_n^0 
~~, 
\end{equation}
which, in the graphical representation, reads:
\begin{equation}
\int {d_H(g_e)}
\begin{array}{c}\setlength{\unitlength}{1 pt}
\begin{picture}(40,40)
    \put(22,2){$\scriptstyle n$}
    \put(0,10){\framebox(40,2){}} \put(20,0){\line(0,1){10}}
    \put(5,12){\line(0,1){3}}\put(5,25){\line(0,1){3}}
    \put(0,15){\framebox(10,10){$\scriptstyle g_e$}}
    \put(12,18){$\cdots$}
    \put(35,12){\line(0,1){3}}\put(35,25){\line(0,1){3}}
    \put(30,15){\framebox(10,10){$\scriptstyle g_e$}}
    \put(0,28){\framebox(40,2){}} \put(20,30){\line(0,1){10}}
    \put(22,34){$\scriptstyle n$}
\end{picture}\end{array}
= \delta_n^0  
~~.
\label{ColourAverage}
\end{equation}
Two representations of $G$ are associated to each edge of  
$\overline{\gamma}$, one of color $n_s$
(eventually 0) from the spin-network $s$ and one of color 
$n_{s'}$ from the spin network $s'$. Using equations
(\ref{eq:REC}) and (\ref{ColourAverage}) one has 
\begin{eqnarray}
\fl
&&  \quad \int {d_H(g_e)} 
\begin{array}{c}\setlength{\unitlength}{1 pt}
\begin{picture}(40,40)
    \put(22,2){$\scriptstyle n_s$}
    \put(0,10){\framebox(40,2){}} \put(20,0){\line(0,1){10}}
    \put(5,12){\line(0,1){3}}\put(5,25){\line(0,1){3}}
    \put(0,15){\framebox(10,10){$\scriptstyle g_e$}}
    \put(12,18){$\cdots$}
    \put(35,12){\line(0,1){3}}\put(35,25){\line(0,1){3}}
    \put(30,15){\framebox(10,10){$\scriptstyle g_e$}}
    \put(0,28){\framebox(40,2){}} \put(20,30){\line(0,1){10}}
    \put(22,34){$\scriptstyle n_s$}
\end{picture}\end{array}
\begin{array}{c}\setlength{\unitlength}{1 pt}
\begin{picture}(40,40)
    \put(22,2){$\scriptstyle n_{s'}$}
    \put(0,10){\framebox(40,2){}} \put(20,0){\line(0,1){10}}
    \put(5,12){\line(0,1){3}}\put(5,25){\line(0,1){3}}
    \put(0,15){\framebox(10,10){$\scriptstyle g_e$}}
    \put(12,18){$\cdots$}
    \put(35,12){\line(0,1){3}}\put(35,25){\line(0,1){3}}
    \put(30,15){\framebox(10,10){$\scriptstyle g_e$}}
    \put(0,28){\framebox(40,2){}} \put(20,30){\line(0,1){10}}
    \put(22,34){$\scriptstyle n_{s'}$}
\end{picture}\end{array}
= \label{EdgeAverage} \\
\fl &&  \quad \qquad ~
= \int {d_H(g_e)} \sum_{k=| n_s - n_{s'}|}^{n_s + n_{s'}}
\left\{\begin{array}{ccc} n_s    & n_s    & k \\
                          n_{s'} & n_{s'} & 0  
\end{array}\right\}
\begin{array}{c}\setlength{\unitlength}{1 pt}
\begin{picture}(40,60)
    \put(20,10){\line(-1,-1){10}}\put(4,2){$\scriptstyle n_{s}$}
    \put(20,10){\line( 1,-1){10}}\put(32,2){$\scriptstyle n_{s'}$}
    \put(20,10){\circle*{3}}
    \put(22,12){$\scriptstyle k$}
    \put(0,20){\framebox(40,2){}} \put(20,10){\line(0,1){10}}
    \put(5,22){\line(0,1){3}}\put(5,35){\line(0,1){3}}
    \put(0,25){\framebox(10,10){$\scriptstyle g_e$}}
    \put(12,28){$\cdots$}
    \put(35,22){\line(0,1){3}}\put(35,35){\line(0,1){3}}
    \put(30,25){\framebox(10,10){$\scriptstyle g_e$}}
    \put(0,38){\framebox(40,2){}} \put(20,40){\line(0,1){10}}
    \put(22,44){$\scriptstyle k$}
    \put(20,50){\line(-1, 1){10}}\put(4,52){$\scriptstyle n_{s}$}
    \put(20,50){\line( 1, 1){10}}\put(32,52){$\scriptstyle n_{s'}$}
    \put(20,50){\circle*{3}}
\end{picture}\end{array}
=\frac{\delta_{n_s}^{n_{s'}}}{\Delta_{n_s}} ~~
\begin{array}{c}\setlength{\unitlength}{1 pt}
\begin{picture}(30,50)
    \put(15,18){\oval(20,10)[t]}
    \put(15,32){\oval(20,10)[b]}
    \put(5,8){\line(0,1){10}}\put(5,32){\line(0,1){10}}
    \put(25,8){\line(0,1){10}}\put(25,32){\line(0,1){10}}
    \put(3,2){$\scriptstyle n_{s}$}\put(3,44){$\scriptstyle n_{s}$}
    \put(23,2){$\scriptstyle n_{s'}$}\put(23,44){$\scriptstyle n_{s'}$}
\end{picture}\end{array}
~~~~.
\nonumber
\end{eqnarray}
This formula shows that the scalar product of two spin-network
states is different from zero only if they have exactly the
same edges and the same coloring
(indeed they also have the same number of vertices and all
the vertices have the same valence).  The implication of
equation (\ref{EdgeAverage}) is even deeper, since it reduces the
computation of the scalar product to the determination of traces
of Clebsh-Gordon coefficients (the recoupling of the vertices),
i.e., to the chromatic evaluation of the corresponding
vertices. 
Indeed, we have found that the scalar product of two
spin-network states is given by:
\begin{equation}
\fl \qquad
  \langle{s},{s'}\rangle 
= \int d\mu_0({\cal A})
  \overline{{\cal T}_{\gamma,\vec{\pi},\vec{c}}[A]}
            {\cal T}_{\gamma',\vec{\pi}',\vec{c}'}[A] 
= \delta_{\gamma,\gamma'} \delta_{\vec{\pi},\vec{\pi}'}
 \prod_{e\in {\cal E}_s} \frac{1}{\Delta_{n_e}} 
 \prod_{i\in {\cal V}_s} \langle V_i , V'_i \rangle
\label{eq:normSN} 
\end{equation}
($\langle V_i , V'_i \rangle$ is the chromatic evaluation 
obtained gluing the vertex $V_i$ and $V_i'$). 
From this expression, it follow that the norm of a spin network 
state $s$ is given by:
\begin{eqnarray}
N[{\gamma,\vec{\pi},\vec{c}}] &=& \sqrt{\langle {s},{s} \rangle}
= \sqrt{ \prod_{i\in{\bar{V}}} \prod_{e\in {\bar{\cal E}}}
         \frac{\theta(a_i,b_i,c_i)}{\Delta_{p_e}} 
       }
\end{eqnarray}
where the product is extended to all edges (including the
virtual ones, i.e., those that come from the $3$-valent 
decomposition) and to all the three-vertices ($a_i,b_i,c_i$
denote the integers that label the three representations in the
three-vertex $i$).  This result, differing from $1$, is the
consequence of the particular normalization chosen
for the fundamental representations and for the
Clebsh-Gordon coefficients; this representation is characterized by
the fact that there are not square root in 
the expressions for the $3n$-$J$ Wigner coefficients. 
It is immediate to note that this is the same expression
found in the loop-representation,
equation (8.7) of  \cite{DePietri96}. 
We will return on this point in section VIII.


\section{Quantization and the operators regularization}

The quantization procedure amounts to choosing an Hilbert space
and a realization of the Poisson algebra of the observables
in terms of Hermitian operators. In the connection representation
the Hilbert space 
$L^2[{\overline{{\cal A}/{\cal G}}},d\mu_0(A)]$
is chosen and the operators realizing the Poisson algebra
are
\begin{eqnarray}
\hat{A}_a^i(x) \cdot f_\gamma(A) &=& {A}_a^i(x) f_\gamma(A) 
\\
\hat{\tilde{E}}{}^a_i(x) \cdot f_\gamma(A) &=&
 - {\rm i} l_0^2 ~\frac{\delta}{\delta A_a^i(x)} f_\gamma(A) ~~.
\end{eqnarray}
The first task that one has to consider in this construction
is the definition of a regularization procedure for the operator
valued distribution $\hat{\tilde{E}}{}^a_i(x)$.  
From the early work
on loop quantum gravity \cite{LOOP} it was realized that in
order to regularize expressions involving this operator it is
necessary to smear $\tilde{E}^a_i(x)$ over a surface
$\Sigma$, and a different regularization surface $\Sigma$  
for any distinct component of the $\tilde{E}^a_i(x)$ operator
has to be chosen.  Supposing one wants to
regularize the components $c_a(x) \tilde{E}^a_i(x)$ of the
$\tilde{E}^a_i(x)$ operator at the point $x$,
one has to consider a
particular two dimensional embedding $\Sigma$ in $M^3$, with
$x\in\Sigma$, $x^a=z^a(\sigma_x)$, such that  $c_a(x) = n_a(z(\sigma_x))$. 
The symbol $\sigma^u=(\sigma^1,\sigma^2)$, ($u,v=1,2$) 
denotes a coordinate system over $\Sigma$ 
($S:\Sigma \longrightarrow M^3, \sigma^u \longrightarrow
~z^a(\sigma)$) and 
\begin{equation}
n_a(z(\sigma)) = \frac{1}{2}\epsilon^{uv} \epsilon_{abc} 
       \frac{\partial z^b}{\partial \sigma^u}
       \frac{\partial z^c}{\partial \sigma^v},
\label{eq:DEFnormal}
\end{equation}
denotes the normal one-form of the embedding $\Sigma$.
Following \cite{Ashtekar96}, the regularization over a 
surface $\Sigma$ will be based on a family of functions, 
dependent on a parameter $\epsilon$, such that
\begin{equation}
  \lim_{\epsilon\rightarrow 0} \int_\Sigma\!d^2\sigma'~
      f_\epsilon(\sigma,\sigma') ~g(\sigma')
  = ~g(\sigma)
\end{equation} 
The smeared version of $n_a \tilde{E}^a_i(x)$ is
indeed  defined as:
\begin{equation}
[c_a \tilde{E}^a_i]_f (x):=  \int_\Sigma\!d^2\sigma
             f_\epsilon(\sigma_x,\sigma) 
             n_a(\sigma) \tilde{E}^a_i(z(\sigma))\, ,
\end{equation}
so that, when $\epsilon$ goes to zero, $[c_a
\tilde{E}^a_i]_f$ goes to the $c_a \tilde{E}^a_i(x)$ 
operator.  This
point-splitting strategy provides a regularized
expression for the $c_a(x) \tilde{E}^a_i(x)$.  Using the
terminology and the graphical representation of the previous
sections, the regularized version of the $c_a(x)\tilde{E}^a_i(x)$ 
operator is given by:
\begin{equation}
\fl \qquad
[c_a\hat{E}^a_i]_f (x) \cdot \Psi_\gamma(A)
    = - {\rm i} l^2_0 \sum_{i=1}^N \! 
        \int_\Sigma d^2\sigma f_\epsilon(\sigma_x,\sigma)
      \left.\frac{\delta g_{e_I}{}^{A}_{~B}}{\delta A_a^i(y)}
      \right|_{y=y(\sigma)} \!\!\!\!  \!\! n_a(\sigma)
      \frac{\partial \Psi_\gamma}{
           \partial g_{e_I}{}^{A}_{~B}}
\label{eq:regE1} 
\end{equation}
where
$$
\frac{\delta g_{e_I}{}}{\delta A_a^i(y)}
  = - \int_0^1 \!\!\! d\lambda \dot{e}^a_i(\lambda)~
       \delta^3\![{e}^a_i(\lambda),y]~
       g_{e_I}[1,\lambda] \tau_i g_{e_I}[\lambda,0]
\label{eq:regE2}
~.
$$
Now, without any loss of generality, it is possible to assume
that the graph $\gamma$, defining the state $\Psi_\gamma$, has
intersections with $\Sigma$ only at its vertices\footnote{
If this is not the case, we can always consider $\Psi_\gamma$
as a cylindrical function $\Psi_{\gamma^\prime}$ of a larger
graph ${\gamma^\prime}$ ($\gamma \subset {\gamma^\prime}$)
that has intersections with $\Sigma$ only at its vertices.
}. Under this
assumption it is possible to perform the integration
in (\ref{eq:regE1}) and one obtains
\begin{equation}
[c_a\hat{E}^a_i]_f (x) \cdot \Psi_\gamma(A)
 = \frac{{\rm i} l^2_0}{2} \sum_{I=1}^N 
   f_\epsilon(\sigma_x,v_k) \tilde{X}^i_{v_k e_I}
   \cdot \Psi_\gamma(A)
\label{eq:E1}
\end{equation}
where
$$
 \tilde{X}^{i}_{v_k e_I} \cdot \Psi_\gamma(A)
    = \left\{\begin{array}{lcl}
        {\displaystyle
        {\varepsilon}_I~ (g_{e_I} \tau_i ) {}^{A}_{~B}
        \frac{\partial \Psi_\gamma}{
              \partial g_{e_I}{}^{A}_{~B}}   } 
        &\quad&\mbox{if }v_k=e_I(0) \\[4 mm] 
        {\displaystyle
        {\varepsilon}_I~ (\tau_i g_{e_I}) {}^{A}_{~B}
        \frac{\partial \Psi_\gamma}{
              \partial g_{e_I}{}^{A}_{~B}}    }
        &\quad&\mbox{if }v_k=e_I(1)  \\[4 mm]
        0 &   &\mbox{otherwise}
        \end{array}\right.
$$
and
$$
{\varepsilon}_I = 
\left\{\begin{array}{lcl}
   0  &\qquad&\mbox{if $\dot{e}_I$ is tangent to $\Sigma$} \\
   +1 &\qquad&\mbox{if $\dot{e}_I$ is directed as $n_a$} \\
   -1 &\qquad&\mbox{if $\dot{e}_I$ is directed opposite to $n_a$.}
\end{array}\right.
$$
We finally remark some facts about the $\tilde{X}^{i}_{v_k e_I}$
vertices operators. 
In order to compute ${\varepsilon}_I$, in the case 
when $\dot{e}_I$ is tangent to $\Sigma$, additional 
regularization is needed, as shown in 
reference \cite{Ashtekar96,Lehner96}.
The relation between these vertices operator and the 
related ones used by Ashtekar and Lewandowski \cite{Ashtekar96}
is $\tilde{X}^{i}_{v_k e_I} = \chi_I {X}^{i}_{I}$.

Considering the graphical representation of a cylindrical function
it is possible to  write the action of the $\tilde{X}^{i}_{v_k e_I}$
vertices operators in a graphical way. 
Let us consider a planar representation in which the
regularizing surface $\Sigma$ is represented by a
horizontal line in the page (the surface in which we 
construct the representation) and the direction of $n_a$ is
directed from the bottom to the top of the page. With this
conventional choice, the four possible cases of the action of
the operators (\ref{eq:E1}) are:
\begin{eqnarray}
\tilde{X}^{i}_{v_k e_I} ~\cdot
~\begin{array}{c}\setlength{\unitlength}{1 pt}
 \begin{picture}(30,40)
    \put(0,0){\framebox(30,10){$\scriptstyle v_k$}}
    \put(15,10){\line(0,1){10}}\put(15,30){\line(0,1){10}}
    \put(10,20){\framebox(10,10){$\scriptstyle g_e$}}
\end{picture}\end{array} 
&=& 
\begin{array}{c}\setlength{\unitlength}{1 pt}
\begin{picture}(40,50)
    \put(0,0){\framebox(30,10){$\scriptstyle v_k$}}
    \put(15,10){\line(0,1){20}}\put(15,40){\line(0,1){10}}
    \put(10,30){\framebox(10,10){$\scriptstyle g_e$}}
    \put(37,16){\line(0,-1){2}}\put(37,24){\line(0,1){2}}
    \put(37,20){\circle{8}} \put(35,17){$\scriptstyle i$}
    \put(37,20){\oval(24,12)[l]}
    \put(18,22){$\scriptstyle 2$}
    \put(15,20){\line(1,0){10}}
    \put(15,20){\circle*{3}}\put(25,20){\circle*{3}}
\end{picture}\end{array} \\
\tilde{X}^{i}_{v_k e_I} ~\cdot
~\begin{array}{c}\setlength{\unitlength}{1 pt}
 \begin{picture}(30,40)
    \put( 0,30){\framebox(30,10){$\scriptstyle v_k$}}
    \put(15, 0){\line(0,1){10}}
    \put(15,20){\line(0,1){10}}
    \put(10,10){\framebox(10,10){$\scriptstyle g_e$}}
\end{picture}\end{array} &=& 
\begin{array}{c}\setlength{\unitlength}{1 pt}
\begin{picture}(40,50)
    \put( 0,40){\framebox(30,10){$\scriptstyle v_k$}}
    \put(15, 0){\line(0,1){10}}
    \put(15,20){\line(0,1){20}}
    \put(10,10){\framebox(10,10){$\scriptstyle g_e$}}
    \put(37,26){\line(0,-1){2}}
    \put(37,34){\line(0,1){2}}
    \put(37,30){\circle{8}} 
    \put(35,27){$\scriptstyle i$}
    \put(37,30){\oval(24,12)[l]}
    \put(18,32){$\scriptstyle 2$}
    \put(15,30){\line(1,0){10}}
    \put(15,30){\circle*{3}}\put(25,30){\circle*{3}}
\end{picture}\end{array} 
~~~~~,
\end{eqnarray}
when the orientation of the edge $e_I$ is from the 
bottom to the top of the page, and
\begin{eqnarray}
\tilde{X}^{i}_{v_k e_I} ~\cdot
~\begin{array}{c}\setlength{\unitlength}{1 pt}   
 \begin{picture}(40,40)
    \put(10,0){\framebox(30,10){$\scriptstyle v_k$}}
    \put(25,10){\line(0,1){20}}
    \put( 5,20){\line(0,1){20}}
    \put(10,20){\oval(10,8)[b]}
    \put(20,30){\oval(10,8)[t]}
    \put(10,20){\framebox(10,10){$\scriptstyle g_e$}}
 \end{picture}\end{array}
&=& - ~~
 \begin{array}{c}\setlength{\unitlength}{1 pt}
 \begin{picture}(50,50)
    \put(10,0){\framebox(30,10){$\scriptstyle v_k$}}
    \put(25,10){\line(0,1){20}}
    \put( 5,20){\line(0,1){30}}
    \put(10,20){\oval(10,8)[b]}
    \put(20,30){\oval(10,28)[t]}
    \put(10,20){\framebox(10,10){$\scriptstyle g_e$}}
    \put(47,31){\line(0,-1){2}}
    \put(47,39){\line(0,1){2}}
    \put(47,35){\circle{8}} 
    \put(45,32){$\scriptstyle i$}
    \put(47,35){\oval(24,12)[l]}
    \put(28,37){$\scriptstyle 2$}
    \put(15,35){\line(1,0){8}}
    \put(27,35){\line(1,0){8}}
    \put(15,35){\circle*{3}}
    \put(35,35){\circle*{3}}
\end{picture}\end{array}
=\begin{array}{c}\setlength{\unitlength}{1 pt}
 \begin{picture}(50,50)
    \put(10,0){\framebox(30,10){$\scriptstyle v_k$}}
    \put(25,10){\line(0,1){20}}
    \put( 5,20){\line(0,1){30}}
    \put(10,20){\oval(10,8)[b]}
    \put(20,30){\oval(10,28)[t]}
    \put(10,20){\framebox(10,10){$\scriptstyle g_e$}}
    \put(47,16){\line(0,-1){2}}
    \put(47,24){\line(0,1){2}}
    \put(47,20){\circle{8}} 
    \put(45,17){$\scriptstyle i$}
    \put(47,20){\oval(24,12)[l]}
    \put(28,22){$\scriptstyle 2$}
    \put(25,20){\line(1,0){10}}
    \put(25,20){\circle*{3}}\put(35,20){\circle*{3}}
\end{picture}\end{array} 
\\
\tilde{X}^{i}_{v_k e_I} ~\cdot
~\begin{array}{c}\setlength{\unitlength}{1 pt}  
 \begin{picture}(40,40)
    \put( 0,30){\framebox(30,10){$\scriptstyle v_k$}}
    \put(25, 0){\line(0,1){20}}
    \put( 5,10){\line(0,1){20}}
    \put(10,10){\oval(10,8)[b]}
    \put(20,20){\oval(10,8)[t]}
    \put(10,10){\framebox(10,10){$\scriptstyle g_e$}}
\end{picture}\end{array} 
&=&  - ~~
\begin{array}{c}\setlength{\unitlength}{1 pt}
\begin{picture}(50,50)
    \put( 0,40){\framebox(30,10){$\scriptstyle v_k$}}
    \put(25, 0){\line(0,1){30}}
    \put( 5,20){\line(0,1){20}}
    \put(10,20){\oval(10,28)[b]}
    \put(20,30){\oval(10,8)[t]}
    \put(10,20){\framebox(10,10){$\scriptstyle g_e$}}
    \put(47,11){\line(0,-1){2}}
    \put(47,19){\line(0,1){2}}
    \put(47,15){\circle{8}} 
    \put(45,12){$\scriptstyle i$}
    \put(47,15){\oval(24,12)[l]}
    \put(28,17){$\scriptstyle 2$}
    \put(15,15){\line(1,0){8}}
    \put(27,15){\line(1,0){8}}
    \put(15,15){\circle*{3}}
    \put(35,15){\circle*{3}}
\end{picture}\end{array} 
=\begin{array}{c}\setlength{\unitlength}{1 pt}
 \begin{picture}(50,50)
    \put( 0,40){\framebox(30,10){$\scriptstyle v_k$}}
    \put(25, 0){\line(0,1){20}}
    \put( 5,10){\line(0,1){30}}
    \put(10,10){\oval(10,8)[b]}
    \put(20,20){\oval(10,8)[t]}
    \put(10,10){\framebox(10,10){$\scriptstyle g_e$}}
    \put(47,26){\line(0,-1){2}}
    \put(47,34){\line(0,1){2}}
    \put(47,30){\circle{8}} 
    \put(45,27){$\scriptstyle i$}
    \put(47,30){\oval(24,12)[l]}
    \put(28,32){$\scriptstyle 2$}
    \put( 5,30){\line(1,0){30}}
    \put( 5,30){\circle*{3}}
    \put(35,30){\circle*{3}}
\end{picture}\end{array} 
\end{eqnarray}
in the other cases.

From the previous formulas it is immediately deduced 
that the graphical expression  of the 
operators $\tilde{X}^{i}_{v_k e_I}$ is independent of
the orientation of the edges of 
the graph (it was already noted that the graphical
representation is independent of the orientation of the edges of
$\gamma$). The action of  $\tilde{X}^{i}_{v_k e_I}$ depends only on
the structure of the contractor associated to the vertex $v_k$:
\begin{equation}
\tilde{X}^{i}_{v_k e_I} ~\cdot
\begin{array}{c}\setlength{\unitlength}{1 pt}
\begin{picture}(50,60)
    \put(10,20){\framebox(30,10){$\scriptstyle c_k$}}
    \put(15,20){\line(0,-1){15}}
    \put(20,10){$...$}
    \put(35,20){\line(0,-1){15}}
    \put(15,30){\line(0,1){15}}
    \put(25,30){\line(0,1){15}}
    \put(22,46){$\scriptstyle n_I$}
    \put(35,30){\line(0,1){15}}
\end{picture}\end{array}
= n_I
\begin{array}{c}\setlength{\unitlength}{1 pt}
\begin{picture}(50,60)
    \put(10,20){\framebox(30,10){$\scriptstyle c_k$}}
    \put(15,20){\line(0,-1){15}}
    \put(20,10){$...$}
    \put(35,20){\line(0,-1){15}}
    \put(15,30){\line(0,1){15}}
    \put(25,30){\line(0,1){15}}
    \put(22,46){$\scriptstyle n_I$}
    \put(27,32){$\scriptstyle n_I$}
    \put(35,30){\line(0,1){15}}
    \put(57,36){\line(0,-1){2}}
    \put(57,44){\line(0,1){2}}
    \put(57,40){\circle{8}} 
    \put(55,37){$\scriptstyle i$}
    \put(57,40){\oval(24,12)[l]}
    \put(38,42){$\scriptstyle 2$}
    \put(25,40){\line(1,0){8}}
    \put(37,40){\line(1,0){8}}
    \put(25,40){\circle*{3}}
    \put(45,40){\circle*{3}}
\end{picture}\end{array}
~~~~~.
\label{eq:XasREC}
\end{equation}
Using equations (\ref{eq:2tau}) and (\ref{eq:3tau})
it is easy to compute the action of the
two gauge invariant vertices operators 
\begin{eqnarray}
\sum_{i} \tilde{X}^{i}_{e_J}\tilde{X}^{i}_{e_I} ~\cdot
\begin{array}{c}\setlength{\unitlength}{1 pt}
\begin{picture}(30,60)
    \put( 0,10){\framebox(30,10){$\scriptstyle c_k$}}
    \put( 5,10){\line(0,-1){5}}
    \put(10, 5){$...$}
    \put(25,10){\line(0,-1){5}}
    \put( 5,20){\line(0,1){25}}
    \put(15,20){\line(0,1){25}}
    \put(12,46){$\scriptstyle n_J$}
    \put(25,20){\line(0,1){25}}
    \put(22,46){$\scriptstyle n_I$}
\end{picture}\end{array}
&=& n_I n_J \sum_i
\begin{array}{c}\setlength{\unitlength}{1 pt}
\begin{picture}(50,60)
    \put( 0,10){\framebox(30,10){$\scriptstyle c_k$}}
    \put( 5,10){\line(0,-1){5}}
    \put(10, 5){$...$}
    \put(25,10){\line(0,-1){5}}
    \put( 5,20){\line(0,1){30}}
    \put(15,20){\line(0,1){30}}
    \put(12,52){$\scriptstyle n_J$}
    \put(25,20){\line(0,1){30}}
    \put(22,52){$\scriptstyle n_I$}
    \put(47,36){\line(0,-1){2}}
    \put(47,44){\line(0,1){2}}
    \put(47,40){\circle{8}} 
    \put(45,37){$\scriptstyle i$}
    \put(47,40){\oval(24,12)[l]}
    \put(28,42){$\scriptstyle 2$}
    \put(15,40){\line(1,0){8}}
    \put(27,40){\line(1,0){8}}
    \put(15,40){\circle*{3}}
    \put(35,40){\circle*{3}}
    \put(47,21){\line(0,-1){2}}
    \put(47,29){\line(0,1){2}}
    \put(47,25){\circle{8}} 
    \put(45,22){$\scriptstyle i$}
    \put(47,25){\oval(24,12)[l]}
    \put(28,27){$\scriptstyle 2$}
    \put(25,25){\line(1,0){10}}
    \put(25,25){\circle*{3}}
    \put(35,25){\circle*{3}}
\end{picture}\end{array}
= \frac{n_I n_J}{2} 
\begin{array}{c}\setlength{\unitlength}{1 pt}
\begin{picture}(50,60)
    \put( 0,10){\framebox(30,10){$\scriptstyle c_k$}}
    \put( 5,10){\line(0,-1){5}}
    \put(10, 5){$...$}
    \put(25,10){\line(0,-1){5}}
    \put( 5,20){\line(0,1){30}}
    \put(15,20){\line(0,1){30}}
    \put(12,52){$\scriptstyle n_J$}
    \put(25,20){\line(0,1){30}}
    \put(22,52){$\scriptstyle n_I$}
    \put(28,42){$\scriptstyle 2$}
    \put(15,40){\line(1,0){8}}
    \put(27,40){\line(1,0){8}}
    \put(15,40){\circle*{3}}
    \put(35,40){\circle*{3}}
    \put(28,27){$\scriptstyle 2$}
    \put(25,25){\line(1,0){10}}
    \put(25,25){\circle*{3}}
    \put(35,25){\circle*{3}}
    \put(35,32.5){\oval(10,15)[r]}
\end{picture}\end{array}
\label{eq:XiXi}
\end{eqnarray}
and,
\begin{equation}
\sum_{i,j,k}\epsilon_{ijk} 
      \tilde{X}^{i}_{e_I}\tilde{X}^{j}_{e_J}\tilde{X}^{k}_{e_K} ~\cdot
\!\!
\begin{array}{c}\setlength{\unitlength}{1 pt}
\begin{picture}(30,60)
    \put( 0,10){\framebox(30,10){$\scriptstyle c_k$}}
    \put( 5,10){\line(0,-1){5}}
    \put(10, 5){$...$}
    \put(25,10){\line(0,-1){5}}
    \put( 5,20){\line(0,1){35}}
    \put( 0,56){$\scriptstyle n_K$}
    \put(15,20){\line(0,1){35}}
    \put(12,56){$\scriptstyle n_J$}
    \put(25,20){\line(0,1){35}}
    \put(22,56){$\scriptstyle n_I$}
\end{picture}\end{array}
= \frac{n_I n_J n_K}{2} \! 
\begin{array}{c}\setlength{\unitlength}{1 pt}
\begin{picture}(50,60)
    \put( 0,10){\framebox(30,10){$\scriptstyle c_k$}}
    \put( 5,10){\line(0,-1){5}}
    \put(10, 5){$...$}
    \put(25,10){\line(0,-1){5}}
    \put( 5,20){\line(0,1){35}}
    \put( 0,56){$\scriptstyle n_K$}
    \put(15,20){\line(0,1){35}}
    \put(12,56){$\scriptstyle n_J$}
    \put(25,20){\line(0,1){35}}
    \put(22,56){$\scriptstyle n_I$}
    \put(28,47){$\scriptstyle 2$}
    \put( 5,45){\line(1,0){8}}
    \put(17,45){\line(1,0){6}}
    \put(27,45){\line(1,0){8}}
    \put( 5,45){\circle*{3}}
    \put(28,37){$\scriptstyle 2$}
    \put(15,35){\line(1,0){8}}
    \put(27,35){\line(1,0){8}}
    \put(15,35){\circle*{3}}
    \put(40,35){\circle*{3}}
    \put(28,27){$\scriptstyle 2$}
    \put(25,25){\line(1,0){10}}
    \put(25,25){\circle*{3}}
    \put(35,35){\oval(10,20)[r]}
    \put(35,35){\line(1,0){5}}
\end{picture}\end{array}
~~.
\label{eq:XiXjXk}
\end{equation}


\section{The binor representation of the 
         area operator}

Two interesting operators for the construction of a 
quantum theory of gravity are the operators associated 
to the classical expression of the area and volume.
The particular interest for these two operators
comes from the suggestion that they will be
true observables in a complete quantum 
theory of gravity coupled to matter \cite{GeomQG}.
We now discuss, in the context of the binor graphical
representation, the operator associated to the classical
area of a surface $\Sigma$ in the connection 
representation of canonical quantum gravity.
We essentially repeat the derivation of the spectrum
of \cite{Ashtekar96}.  As a consequence, it will be
shown that the mathematical steps required in the
computation are exactly the same as those of the loop
representation \cite{DePietri96,Rovelli95a,Lehner96}.

A surface $\Sigma$ in $M$ is an embedding of a
2-dimensional manifold $\Sigma$, with coordinates
$\sigma^u=(\sigma^1,\sigma^2)$,  into $M$.  
We write $S:\Sigma \longrightarrow M^3, \sigma^u
\longrightarrow ~x^a(\sigma)$. 
Denoting the normal one-form with $n_a$ and the 
induced metric on $\Sigma$ with $g^{\Sigma}_{uv}$,
the classical expression for the area of $\Sigma$ is 
\begin{equation}
A[\Sigma] = \int_\Sigma\!d^2\sigma~\sqrt{\det g^{\Sigma}} 
= \int_\Sigma\! d^2\sigma
~\sqrt{n_a n_b \tilde{E}^{ai} \tilde{E}^{b}_{i}} ~. 
\label{area}
\end{equation}
The quantum area operator will be defined using the 
regularization of the $\tilde{E}^{ai}$ operators defined in the
previous section, and the regularizing surface will be chosen to be
$\Sigma$ itself.  
From equation (\ref{eq:E1}) 
the following expression for the regularized area operator
is obtained:
\begin{eqnarray}
\hat{A}^2[\Sigma] 
   \cdot \Psi_\gamma(A) 
&=& \int_\Sigma d^2\!x \sqrt{\sum_i 
       [n_a\hat{E}^a_i]_f(x)
       [n_a\hat{E}^a_i]_f(x) } 
  \cdot \Psi_\gamma(A)
\label{def:Area} \\
&=&
 \int_\Sigma l_0^2
  d^2\!x~ \sum_{k=1}^{M} f_\epsilon(\sigma_x,v_k)
  \sqrt{ \hat{A}^2[\Sigma,v_k] }
  \cdot \Psi_\gamma(A)
~~.
\nonumber
\end{eqnarray}
Here, $\hat{A}^2[\Sigma,v_k]$ is the recoupling vertex operator 
corresponding to the area contribution due to the vertex
$v_k$ of the spin-network state $\gamma$
\begin{equation}
\hat{A}^2[\Sigma,v_k]  
 \cdot \Psi_\gamma(A)
= \sum_{I,J=1}^N \sum_{i=1}^{3} \frac{-1}{4} 
         \tilde{X}^{i}_{v_k e_I} \tilde{X}^{i}_{v_k e_J}
  \cdot \Psi_\gamma(A) 
~~.
\label{def:AreaVertex}
\end{equation}   
In the previous section it was shown how a generic
cylindrical function can be written in a graphical planar
representation inside $\Gamma_{ex}$ and how the action
of the vertices operator 
$\tilde{X}^{i}_{v_k e_I}$ can be written in 
a graphical way. 
Indeed, given a graph $\gamma$, consider one of its
bidimensional graphical representations. In this
plane, the set of the edges at the vertex $v_k$ 
(that lies over $\Sigma$) are naturally decomposed 
in $3$ distinct subclasses: 
{\rm (i)}   those that lie above $\Sigma$;
{\rm (ii)}  those that lie below $\Sigma$; 
{\rm (iii)} those that lie tangential to $\Sigma$.
From the recoupling theorem it is always possible to parameterize 
the contractor $c_k$ associated to the vertex $v_k$ as
\begin{equation}
\begin{array}{c}\setlength{\unitlength}{1 pt}
\begin{picture}(110,60)
    \put(67, 5){\rule{6pt}{10pt}}
    \put(70, 0){\line(0,1){5}}
    \put(70,15){\line(0,1){5}}
    \put(70,10){$\ldots$}
    \put(92, 0){$\scriptstyle n_{u+d+t}$}
    \put(95,10){${e_{u+d+t}}$}
    \put(87, 5){\rule{6pt}{10pt}}
    \put(90, 0){\line(0,1){5}}
    \put(90,15){\line(0,1){5}}
    \put(40,20){\oval( 60, 4)[rt]}
    \put(40,20){\oval(100,16)[rt]}
    \put( 7, 0){$\scriptstyle n_{u+1}$}
    \put(10,10){${e_{u+1}}$}
    \put( 2, 5){\rule{6pt}{10pt}}
    \put( 5, 0){\line(0,1){5}}
    \put( 5,15){\line(0,1){5}}
    \put(20,-5){$..$}
    \put(37, 0){$\scriptstyle n_{u+d}$}
    \put(40,10){${e_{u+d}}$}
    \put(32, 5){\rule{6pt}{10pt}}
    \put(35, 0){\line(0,1){5}}
    \put(35,15){\line(0,1){5}}
    \put(0,20){\framebox(40,10){$c_k$}}
    \put( 7,52){$\scriptstyle n_{1}$}
    \put(10,40){${e_{1}}$}
    \put( 2,35){\rule{6pt}{10pt}}
    \put( 5,30){\line(0,1){5}}
    \put( 5,45){\line(0,1){5}}
    \put(20,40){$..$}
    \put(37,52){$\scriptstyle n_{u}$}
    \put(40,40){${e_{u}}$}
    \put(32,35){\rule{6pt}{10pt}}
    \put(35,30){\line(0,1){5}}
    \put(35,45){\line(0,1){5}}
\end{picture}\end{array}
= \begin{array}{c}\setlength{\unitlength}{1 pt}
  \begin{picture}(100,90)
    \put(67, 5){\rule{6pt}{10pt}}
    \put(70, 0){\line(0,1){5}}
    \put(70,15){\line(0,1){5}}
    \put(75,10){$..$}
    \put(87, 5){\rule{6pt}{10pt}}
    \put(90, 0){\line(0,1){5}}
    \put(90,15){\line(0,1){5}}
    \put(65,20){\oval( 10,34)[rt]}
    \put(65,20){\oval( 50,46)[rt]}
    \put( 7, 0){$\scriptstyle n_{u+1}$}
    \put(10,10){${e_{u+1}}$}
    \put( 2, 5){\rule{6pt}{10pt}}
    \put( 5, 0){\line(0,1){5}}
    \put( 5,15){\line(0,1){5}}
    \put(20,10){$..$}
    \put(37, 0){$\scriptstyle n_{u+d}$}
    \put(40,10){${e_{u+d}}$}
    \put(32, 5){\rule{6pt}{10pt}}
    \put(35, 0){\line(0,1){5}}
    \put(35,15){\line(0,1){5}}
    \put(0,20){\framebox(40,10){$\scriptstyle c_k^{(d)}$}}
    \put(20,30){\line(0,1){20}}
    \put(20,40){\circle*{3}}
    \put(22,32){$\scriptstyle d$}
    \put(22,42){$\scriptstyle u$}
    \put(35,42){$\scriptstyle t$}
    \put(20,40){\line(1,0){25}}
    \put(0,50){\framebox(40,10){$\scriptstyle c_k^{(u)}$}}
    \put( 7,82){$\scriptstyle n_{1}$}
    \put(10,70){${e_{1}}$}
    \put( 2,65){\rule{6pt}{10pt}}
    \put( 5,60){\line(0,1){5}}
    \put( 5,75){\line(0,1){5}}
    \put(20,70){$..$}
    \put(37,82){$\scriptstyle n_{u}$}
    \put(40,70){${e_{u}}$}
    \put(32,65){\rule{6pt}{10pt}}
    \put(35,60){\line(0,1){5}}
    \put(35,75){\line(0,1){5}}
    \put(45,32){\framebox(20,16){$\scriptstyle c_k^{(t)}$}}
\end{picture}\end{array}
\label{eq:vertexDEC}
\end{equation} 
From the identity\footnote{This identity 
can be easily obtained using the definition 
of the contractors (\ref{eq:7}) and the expansion
of the three-vertices (\ref{eq:6}). For a deduction based only 
on the recoupling theorem see \cite{Lehner96}.} 
\begin{equation}
\sum_{k=1}^{u} ~n_k~
\begin{array}{c}\setlength{\unitlength}{1 pt}
\begin{picture}(85,60)
    \put(30, 0){\line(0,1){10}}
    \put(32, 0){$\scriptstyle u$}
    \put(0,10){\framebox(60,10){$c_k$}}
    \put( 7,52){$\scriptstyle n_{1}$}
    \put(10,40){${e_{1}}$}
    \put( 2,35){\rule{6pt}{10pt}}
    \put( 5,20){\line(0,1){15}}
    \put( 5,45){\line(0,1){5}}
    \put(20,40){$..$}
    \put(32,52){$\scriptstyle n_{k}$}
    \put(35,40){${e_{k}}$}
    \put(27,35){\rule{6pt}{10pt}}
    \put(30,20){\line(0,1){15}}
    \put(30,45){\line(0,1){5}}
    \put(45,40){$..$}
    \put(57,52){$\scriptstyle n_{u}$}
    \put(60,40){${e_{u}}$}
    \put(52,35){\rule{6pt}{10pt}}
    \put(55,20){\line(0,1){15}}
    \put(55,45){\line(0,1){5}}
    \put(30,30){\circle*{3}}
    \put(30,30){\line(1,0){22}}
    \put(65,30){\line(-1,0){8}}
    \put(77,26){\line(0,-1){2}}
    \put(77,34){\line(0,1){2}}
    \put(77,30){\circle{8}} 
    \put(75,27){$\scriptstyle i$}
    \put(77,30){\oval(24,12)[l]}
    \put(58,24){$\scriptstyle 2$}
    \put(65,30){\circle*{3}}
\end{picture}\end{array}
= ~u~
 \begin{array}{c}\setlength{\unitlength}{1 pt}
 \begin{picture}(80,60)
    \put(30, 0){\line(0,1){20}}
    \put(32, 0){$\scriptstyle u$}
    \put(0,20){\framebox(60,10){$c_k$}}
    \put( 7,52){$\scriptstyle n_{1}$}
    \put(10,40){${e_{1}}$}
    \put( 2,35){\rule{6pt}{10pt}}
    \put( 5,30){\line(0,1){5}}
    \put( 5,45){\line(0,1){5}}
    \put(20,40){$..$}
    \put(32,52){$\scriptstyle n_{k}$}
    \put(30,40){${e_{k}}$}
    \put(27,35){\rule{6pt}{10pt}}
    \put(30,30){\line(0,1){5}}
    \put(30,45){\line(0,1){5}}
    \put(45,40){$..$}
    \put(57,52){$\scriptstyle n_{u}$}
    \put(60,40){${e_{u}}$}
    \put(52,35){\rule{6pt}{10pt}}
    \put(55,30){\line(0,1){5}}
    \put(55,45){\line(0,1){5}}
    \put(30,10){\circle*{3}}
    \put(30,10){\line(1,0){30}}
    \put(72, 6){\line(0,-1){2}}
    \put(72,14){\line(0,1){2}}
    \put(72,10){\circle{8}} 
    \put(70, 7){$\scriptstyle i$}
    \put(72,10){\oval(24,12)[l]}
    \put(50,12){$\scriptstyle 2$}
    \put(60,10){\circle*{3}}
  \end{picture}\end{array}
\end{equation} 
it is possible to  reduce the computation of the area vertex 
operator to the three-valent vertex with only three edges 
(possibly of color $0$) in which one lies above $\Sigma$, one
lies below $\Sigma$ and one is tangent to $\Sigma$.
Indeed, one has to compute its action only on 
the three valent vertex $V_3(d,u,t)$ 
where the line of color $u$ correspond to the edges
above $\Sigma$, the line of color $d$ to the edges
below $\Sigma$ and the line of color $t$ to the edges
tangent to $\Sigma$. Now, from the result of section 
VII one has that the operators $X^{i}_{v_k e_I}$ give 
contributions only when they are applied to an edge 
which is not tangent  to $\Sigma$. 
Using this fact and the identities (\ref{eq:XiXi})
one has the following three contribution to the action of the
$\hat{A}^2[\Sigma,v_k]$ vertex operator
\begin{eqnarray}
\sum_{i=1}^{3} \tilde{X}^{i}_{e_u} \tilde{X}^{i}_{e_u}
\begin{array}{c}\setlength{\unitlength}{1 pt}
\begin{picture}(25,20)
  \put(17, 0){$\scriptstyle {t}$}
  \put(15, 0){\line(0,1){5}}
  \put( 5, 5){\oval(20,10)[rt]}
  \put( 7, 0){$\scriptstyle {d}$}
  \put( 5, 0){\line(0,1){10}}
  \put( 5,10){\circle*{3}}
  \put( 7,16){$\scriptstyle u$}
  \put( 5,10){\line(0,1){10}}
\end{picture}\end{array}
  &=& \frac{u^2}{2}
\begin{array}{c}\setlength{\unitlength}{1 pt}
\begin{picture}(30,40)
    \put(22, 0){$\scriptstyle {t}$}
    \put(10, 0){\oval(20,20)[rt]}
    \put(12, 0){$\scriptstyle {d}$}
    \put(10, 0){\line(0,1){20}}
    \put(10,20){\circle*{3}}
    \put(12,36){$\scriptstyle u$}
    \put(10,20){\line(0,1){20}}
    \put(10,10){\circle*{3}}
    \put(10,25){\oval(20,10)[l]}
    \put(10,30){\circle*{3}}
    \put( 2,22){$\scriptstyle 2$}
    \put(12,12){$\scriptstyle u$}
    \put(12,22){$\scriptstyle u$}
\end{picture}\end{array}
= A^2_u
\begin{array}{c}\setlength{\unitlength}{1 pt}
\begin{picture}(25,20)
  \put(17, 0){$\scriptstyle {t}$}
  \put(15, 0){\line(0,1){5}}
  \put( 5, 5){\oval(20,10)[rt]}
  \put( 7, 0){$\scriptstyle {d}$}
  \put( 5, 0){\line(0,1){10}}
  \put( 5,10){\circle*{3}}
  \put( 7,16){$\scriptstyle u$}
  \put( 5,10){\line(0,1){10}}
\end{picture}\end{array}
\\
\sum_{i=1}^{3} \tilde{X}^{i}_{e_d} \tilde{X}^{i}_{e_d}
\begin{array}{c}\setlength{\unitlength}{1 pt}
\begin{picture}(25,20)
  \put(17, 0){$\scriptstyle {t}$}
  \put(15, 0){\line(0,1){5}}
  \put( 5, 5){\oval(20,10)[rt]}
  \put( 7, 0){$\scriptstyle {d}$}
  \put( 5, 0){\line(0,1){10}}
  \put( 5,10){\circle*{3}}
  \put( 7,16){$\scriptstyle u$}
  \put( 5,10){\line(0,1){10}}
\end{picture}\end{array}
  &=& \frac{d^2}{2}
\begin{array}{c}\setlength{\unitlength}{1 pt}
\begin{picture}(30,40)
    \put(22, 0){$\scriptstyle {t}$}
    \put(20, 0){\line(0,1){10}}
    \put(10,10){\oval(20,40)[rt]}
    \put(12, 0){$\scriptstyle {d}$}
    \put(10, 0){\line(0,1){20}}
    \put(10,20){\circle*{3}}
    \put(12,36){$\scriptstyle u$}
    \put(10,20){\line(0,1){20}}
    \put(10,10){\circle*{3}}
    \put(10,15){\oval(20,10)[l]}
    \put(10,30){\circle*{3}}
    \put( 2,12){$\scriptstyle 2$}
    \put(12,12){$\scriptstyle d$}
    \put(12,22){$\scriptstyle d$}
\end{picture}\end{array}
= A^2_d
\begin{array}{c}\setlength{\unitlength}{1 pt}
\begin{picture}(25,20)
  \put(17, 0){$\scriptstyle {t}$}
  \put(15, 0){\line(0,1){5}}
  \put( 5, 5){\oval(20,10)[rt]}
  \put( 7, 0){$\scriptstyle {d}$}
  \put( 5, 0){\line(0,1){10}}
  \put( 5,10){\circle*{3}}
  \put( 7,16){$\scriptstyle u$}
  \put( 5,10){\line(0,1){10}}
\end{picture}\end{array}
\\
\sum_{i=1}^{3} \tilde{X}^{i}_{e_u} \tilde{X}^{i}_{e_d}
\begin{array}{c}\setlength{\unitlength}{1 pt}
\begin{picture}(25,20)
  \put(17, 0){$\scriptstyle {t}$}
  \put(15, 0){\line(0,1){5}}
  \put( 5, 5){\oval(20,10)[rt]}
  \put( 7, 0){$\scriptstyle {d}$}
  \put( 5, 0){\line(0,1){10}}
  \put( 5,10){\circle*{3}}
  \put( 7,16){$\scriptstyle u$}
  \put( 5,10){\line(0,1){10}}
\end{picture}\end{array}
  &=& \frac{ud}{2}
\begin{array}{c}\setlength{\unitlength}{1 pt}
\begin{picture}(30,40)
    \put(22, 0){$\scriptstyle {t}$}
    \put(20, 0){\line(0,1){10}}
    \put(10,10){\oval(20,20)[rt]}
    \put(12, 0){$\scriptstyle {d}$}
    \put(10, 0){\line(0,1){20}}
    \put(10,20){\circle*{3}}
    \put(12,36){$\scriptstyle u$}
    \put(10,20){\line(0,1){20}}
    \put(10,10){\circle*{3}}
    \put(10,20){\oval(20,20)[l]}
    \put(10,30){\circle*{3}}
    \put( 2,17){$\scriptstyle 2$}
    \put(12,12){$\scriptstyle d$}
    \put(12,22){$\scriptstyle u$}
\end{picture}\end{array}
= A^2_t
\begin{array}{c}\setlength{\unitlength}{1 pt}
\begin{picture}(25,20)
  \put(17, 0){$\scriptstyle {t}$}
  \put(15, 0){\line(0,1){5}}
  \put( 5, 5){\oval(20,10)[rt]}
  \put( 7, 0){$\scriptstyle {d}$}
  \put( 5, 0){\line(0,1){10}}
  \put( 5,10){\circle*{3}}
  \put( 7,16){$\scriptstyle u$}
  \put( 5,10){\line(0,1){10}}
\end{picture}\end{array}
\end{eqnarray}
where the number $A^2_u$, $A^2_d$ and $A^2_t$ can be
calculated using the scalar product in the space of 
the contractors of a vertex defined by the chromatic 
evaluation (see section II). The explicit expression 
for $A^2_u$, $A^2_d$ and $A^2_t$ are
\begin{eqnarray}
A_u^2(u,d,t) &=& \frac{u^2}{2} 
\frac{\begin{array}{c}\setlength{\unitlength}{.5 pt}
     \begin{picture}(40,40)
        \put(18,32){$\scriptstyle 2$}
        \put( 0,15){\line(1,0){40}} \put(18,17){$\scriptstyle u$}
        \put(20,15){\oval(40,30)}   \put(18, 2){$\scriptstyle u$}
        \put( 0,15){\circle*{3}}    \put(40,15){\circle*{3}}
     \end{picture}\end{array}
   }{\begin{array}{c}\setlength{\unitlength}{.5 pt}
     \begin{picture}(35,25)
         \put(15, 0){\line(0,1){10}}\put(20, 0){\line(0,1){10}}
         \put(15, 0){\line(1,0){ 5}}\put(15,10){\line(1,0){ 5}}
         \put(15,25){\line(1,0){5}}  \put(15,15){$\scriptstyle u$}
         \put(15,15){\oval(30,20)[l]}\put(20,15){\oval(30,20)[r]} 
     \end{picture}\end{array}
    }
  = - \frac{u(u+2)}{4}
\\
A_t^2(u,d,t) &=&  \frac{ud}{2} 
\frac{\begin{array}{c}\setlength{\unitlength}{.5 pt}
     \begin{picture}(60,40)
        \put(20,18){$\scriptstyle u$}
        \put(20, 2){$\scriptstyle d$}
        \put( 5,12){$\scriptstyle 2$}
        \put(30, 0){\line(0,1){30}}
        \put(30, 0){\circle*{3}} \put(30,30){\circle*{3}}
        \put(42,32){$\scriptstyle u$}
        \put(30,15){\line(1,0){30}} \put(42,17){$\scriptstyle t$}
        \put(30,15){\oval(60,30)}   \put(42, 2){$\scriptstyle d$}
        \put(30,15){\circle*{3}}    \put(60,15){\circle*{3}}
     \end{picture}\end{array}
   }{\begin{array}{c}\setlength{\unitlength}{.5 pt}
     \begin{picture}(40,40)
        \put(18,32){$\scriptstyle u$}
        \put( 0,15){\line(1,0){40}} \put(18,17){$\scriptstyle t$}
        \put(20,15){\oval(40,30)}   \put(18, 2){$\scriptstyle d$}
        \put( 0,15){\circle*{3}}    \put(40,15){\circle*{3}}
     \end{picture}\end{array}
   } 
 = - \frac{u(u+2)+d(d+2)-t(t+2)}{8}
~~.
\end{eqnarray}
and $A^2_d(u,d,t)= A^2_u(d,u,t)$.
In this way, one has obtained that the vertices
written in the r.h.s.\ of equation (\ref{eq:vertexDEC}) 
are eigenvectors of the  $\hat{A}^2[\Sigma,v_k]$ operator
corresponding to the eigenvalues: 
\begin{eqnarray}
\fl
&& A^2[\Sigma,v_k[u,d,t]] 
    = -\frac{A^2_u + A^2_d + 2 A^2_t}{4}
\\
\fl
&&\qquad = -\frac{1}{4} \bigg[ 
    - \frac{u (u+2)}{4} 
    - \frac{d (d+2)}{4} 
    - 2 \frac{u(u+2)+d(d+2)-t(t+2)}{8}
\bigg]
\nonumber \\
\fl
&&\qquad =  \frac{u (u+2)}{8} 
   + \frac{d (d+2)}{8}
   - \frac{t(t+2)}{16}
 =  \frac{j_u (j_u+1)}{2} 
   + \frac{j_d (j_d+1)}{2}
   - \frac{j_t(j_t+1)}{4}
\nonumber
\end{eqnarray}
The full spectrum of the area operator\footnote{
In this way we have re-derived the spectrum
computed in \cite{Ashtekar96,Lehner96}.
} is indeed
\begin{equation} 
 \hat{A}^2[\Sigma] \cdot
 = l_0^2 \sum_{v_k\in \gamma\cap\Sigma} 
    \sqrt{A^2[\Sigma,v_k[u,d,t]]  }
 \Psi_\gamma(A)
\end{equation}
where $\Psi_\gamma(A)$ is a spin network 
state ${\cal T}_{\gamma,\vec{\pi},\vec{c}}[A]$ in which
all the intersection $\gamma\cap\Sigma$ are vertices of the 
spin-network state and all the vertices $v_k\in \gamma\cap\Sigma$
are decomposed according to the r.h.s.\ of equation 
(\ref{eq:vertexDEC}).


\section{The loop transformation}

As noted in \cite{Rovelli95} the loop representation
${\cal R}_l$ has to be identified with the representation 
$\overline{{\cal R}}_c$ dual to the connection 
representation ${\cal R}_c$. For convenience
we will denote by $\langle d\mu |$ the bra states
of ${\cal R}_c$ that, by definition, are the ket 
states  $| d\mu \rangle$ of $\overline{{\cal R}}_c$.
Clearly, in absence of an inner product, there is
no canonical map between the connection 
representation ${\cal R}_c$ and the dual-representation  
$\overline{{\cal R}}_c$ and, indeed, between the connection
${\cal R}_c$ and the loop ${\cal R}_l$ representation.
In detail we have that any functional of the connection
representation $\psi(A)=\langle A | \psi \rangle$
defines, by double duality, a linear map on the 
state space $\overline{{\cal R}}_c$
\begin{equation}
  \langle \psi | d\mu \rangle 
= \overline{\langle d\mu | \psi \rangle}
= \overline{\int d\mu(A) ~\psi(A) }
~~~.
\end{equation} 
Given a loop state $| \psi_\alpha \rangle$,
which is defined in the connection representation
${\cal R}_c$ by
\begin{equation} 
 \langle A | \psi_\alpha \rangle 
 = {\cal T}[A,\alpha] 
 = - {\rm Tr}[{\cal P} e^{-\int_\alpha dx^a A_a} ] 
~~,
\label{eq:loopDEF}
\end{equation}
it determines a dual state $\langle \psi_\alpha |$
in $\overline{{\cal R}}_c$, via
\begin{equation}
  \langle \psi_\alpha | d\mu \rangle 
= \overline{\langle d\mu | \psi_\alpha \rangle}
= \overline{\int\! d\mu(A) ~{\cal T}[A,\alpha] }
~~,
\end{equation} 
and this bra state must be identified precisely
with the loop state
\begin{equation}
  {}_L\langle \alpha | = \langle \psi_\alpha | ~~.
\end{equation}
In the case that a scalar product is defined in the 
connection representation, it is possible to associate
to any functional $\psi(A)\in {\cal R}_c$ a unique
dual state $\langle d\mu_\psi |$ such that, for 
any $\psi^\prime(A)$,
\begin{equation}
 \langle d\mu_\psi |\psi^\prime \rangle
  = \int\! d\mu_\psi(A) \psi^\prime(A)
  = \int\! d\mu_0(A)~ \psi(A) \psi^\prime(A).
\label{eq:duality}
\end{equation}
From equation (\ref{eq:loopDEF}) and the duality defined by 
equation (\ref{eq:duality}), the definition 
of the loop transform is given by
\begin{equation}
\psi(\alpha) = {}_L\langle \alpha | \psi \rangle 
  = \int\! d\mu_0(A)~\overline{{\cal T}[A,\alpha]} ~\psi(A)
~~.
\end{equation}
In particular, if a scalar product is given in the
connection representation, a scalar product in the loop representation
${\cal R}_l$ is defined by
\begin{equation}
 {}_L\langle \alpha |\beta \rangle{}_L 
=\int d\mu_0(A) {}_L\langle\alpha|A\rangle\langle A|\beta\rangle{}_L
= \int d\mu_0(A) \overline{{\cal T}[A,\alpha]}{\cal T}[A,\beta]
~.
\label{eq:loopSP}
\end{equation}
A direct examination of the definition of the spin-network
states in the loop representation 
(equations (2.12),(2.15) and section V of \cite{DePietri96})
and a comparison with the definition (\ref{eq:defSPINnet}), 
shows that
\begin{equation}
 \langle A | \gamma,\vec{\pi},\vec{c} \rangle
   =  {\cal T}_{\gamma,\vec{\pi},\vec{c}}[A]
~~.
\label{eq:SPINNETloop}
\end{equation}
Using equations (\ref{eq:normSN}), (\ref{eq:loopSP}) and
(\ref{eq:SPINNETloop}) one has that the scalar product induced 
in the loop representation is
\begin{equation}
 \langle{\gamma,\vec{\pi},\vec{c}}|
        {\gamma',\vec{\pi}',\vec{c}'}\rangle
 =  \delta_{\gamma,\gamma'} \delta_{\vec{\pi},\vec{\pi}'}
      N^2[{\gamma,\vec{\pi},\vec{c}}]
\end{equation}
which is exactly the result for the proposed scalar product
in the loop representation of \cite{DePietri96},
section VIII. 


\section{Conclusions} 

In this article a graphical representation for the cylindrical 
functions in $L^2[{\overline{{\cal A}/{\cal G}}},d\mu_0(A)]$ 
and for the spin-network basis has been explicitly constructed.  
Using a particular choice of
irreducible representations of $G$ and of contractors,
a graphical representation for the
regularized operator $[c_a \tilde{E}^a_i]_f (x)$
has been obtained.
Moreover, using these constructions, the equivalence of the area 
operator in the loop and in the connection representation
of Euclidean quantum gravity has been proven.
Finally, it has been shown that the scalar product proposed in 
the loop representation of Ref.\ \cite{DePietri96} is exactly 
the one induced by the loop-transformation of the 
Ashtekar-Lewandowski measure $d\mu_0(A)$.


\ack

I am particularly grateful to Carlo Rovelli for 
valuable criticisms and insight. 
Moreover, I would like to thank: Simonetta Frittelli
and Luis Lehner for having shared with me the results 
contained in Ref.\ \cite{Lehner96} prior to publication; 
Massimo Pauri and Luca Lusanna for their continuous support 
and encouragement during these years.  
This work has been partially supported
by the INFN grant ``Iniziativa specifica FI-41'' (Italy), 
and by the Human Capital and Mobility Program 
``Constrained Dynamical Systems'' (European Union).




\Figures

\end{document}